\documentclass[journal]{IEEEtran}

\IEEEoverridecommandlockouts

\usepackage{cite}
\usepackage{graphicx}
\usepackage[usenames,dvipsnames]{color}
\usepackage{amsmath}
\usepackage{epstopdf}
\usepackage{multicol}
\usepackage{multirow}
\usepackage{booktabs}
\usepackage[hyphenbreaks]{breakurl}
\usepackage[hyphens]{url}
\usepackage[caption=false,font=footnotesize]{subfig}

\begin{document}
	
\title{Radiation Testing of Optical and Semiconductor Components for Radiation-Tolerant LED Luminaires}

\author{Alessandro Floriduz, \IEEEmembership{Member, IEEE}, and James D. Devine, \IEEEmembership{Member, IEEE}%
	\thanks{A. Floriduz and J. D. Devine are with the Electrical Network Projects Section of the Electrical Group, Engineering Department (EN-EL), CERN (European Organization for Nuclear Research), CH-1211 Geneva 23, Switzerland {(email: alessandro.floriduz@alumni.cern, james.dilwyn.devine@cern.ch)}.}
}

\markboth{Radiation Effects on Components and Systems (RADECS) Conference, 16--21 September 2018, Gothenburg, Sweden}%
{Conference on Radiation Effects on Components and Systems (RADECS), 16--21 September 2018, Gothenburg, Sweden}

\maketitle

\begin{abstract}
An irradiation campaign was conducted to provide guidance in the selection of materials and components for the radiation hardening of LED lights for use in CERN accelerator tunnels. This work describes the effects of gamma-rays on commercial-grade borosilicate, fused quartz, polymethylmethacrylate, and polycarbonate samples up to doses of 100~kGy, to qualify their use as optical materials in rad-hard LED-based luminaires. In addition, a Si bridge rectifier and a SiC Junction Barrier Schottky diode for use in power supplies of rad-hard LED lighting systems are tested using 24~GeV/\emph{c} protons. The physical degradation mechanisms are discussed for each element. 

\end{abstract}

% Commentato per abstract.
% Keywords - change from "Index terms" to "Keywords"
\renewcommand\IEEEkeywordsname{Keywords}

\section{Introduction}
\begin{figure}[!t]
	\centering
	\includegraphics[width=0.975\columnwidth]{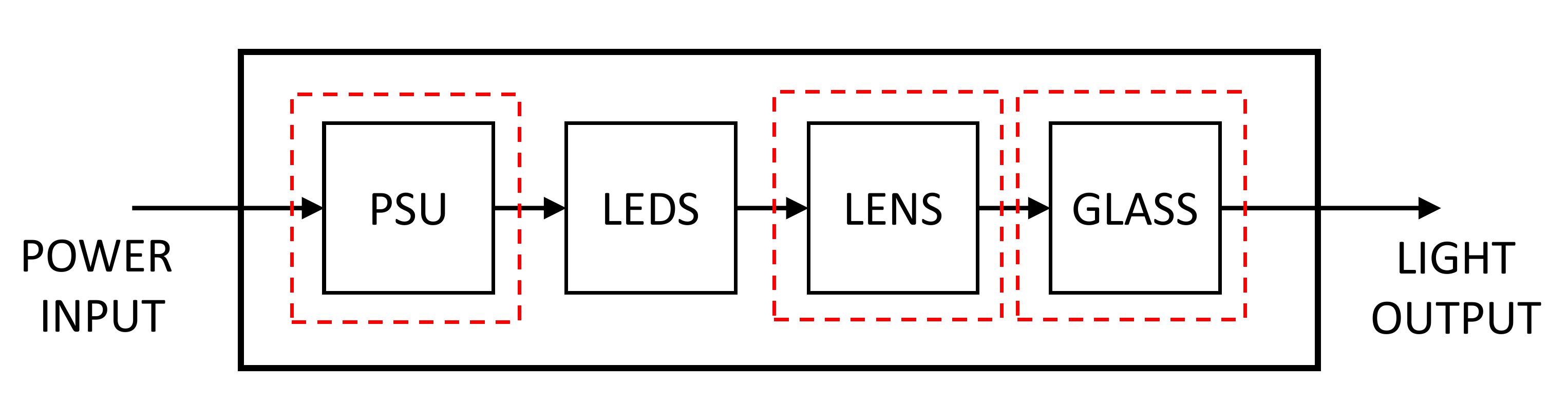}
	\caption{Block diagram model of an LED luminaire, which comprises a power supply unit (PSU) feeding a string of white LEDs; it also includes plastic lenses for improved light distribution and a glass window for protection reasons. The response to radiation of the components highlighted in red is illustrated in the present work.}
	\label{fig:model}
\end{figure}

\IEEEPARstart{G}{ENERAL} lighting in CERN accelerator tunnels and experimental caverns is currently provided by fluorescent tubes with wire-wound ballasts, and the present emergency lighting system comprises a combination of incandescent and low-pressure sodium discharge lamps. However, luminaires using these technologies are now obsolete and cannot be purchased any more on the European market, and thus have to be replaced by LED lights having much higher efficiency. Nevertheless, adoption of LED-based luminaires in radiation environments like high-energy physics accelerator facilities  poses several challenges due to the radiation effects on semiconductor components \cite{devine2016radiation}; indeed, we recall that, on the walls of a typical section of CERN accelerator tunnels, the annual radiation levels can exceed a 1~MeV neutron equivalent (neq) fluence in Si of 5$\times$10$^\text{12}$~n/cm$^\text{2}$ and a dose of 1~kGy \cite{de2015radiation}. %; conversely, traditional lighting systems proved to be very robust against radiation. 
Irradiation of prototype LED lights for radiation environment has been carried out previously, results of which are summarized in \cite{devine2016radiation}. 
Following the encouraging outcome of these preliminary tests, a dedicated irradiation campaign has been performed to study the effects of radiation on each individual component to be included in a radiation-tolerant LED lamp (see Fig. \ref{fig:model}), namely: \emph{i)}~the diodes for the power supply unit (consisting of a simple bridge rectifier), \emph{ii)} the plastic lenses and glass windows, and \emph{iii)} the GaN-based white LEDs. This paper summarizes the results on irradiation tests of the first two components; the behaviour of irradiated high-power white GaN LEDs is available in \cite{floriduz2018modelling}.

% Motivation
In this work, we tested  under $\gamma$-ray irradiation samples of borosilicate (BS), fused quartz (FQ), polymethylmethacrylate (PMMA) and polycarbonate (PC). We chose to study these materials as they are commonly used in commercial luminaires: glass (BS or FQ) for protective windows, and plastic (PMMA or PC) for secondary optics. Reports on the irradiation of these materials already exist, but tests under the same irradiation and annealing conditions were required to better compare the amount of radiation damage induced to the various materials, so as to identify the most resistant ones. %The motivation for our test is to assess the behaviour of these material under the same irradiation and annealing conditions, to better compare identify the best candidate material.

In addition, a Si bridge rectifier and a SiC Junction Barrier Schottky (JBS) diode have been tested; irradiation was done using 24~GeV/\emph{c} protons instead of $\gamma$-rays, displacement damage being the dominant degradation mechanism in diodes. A Si diode bridge sharing the same technology as the selected one had already been tested against irradiation up to a 1~MeV neq fluence of 8$\times$10$^\text{13}$~n/cm$^\text{2}$ \cite{bager2002lhcb}; in the present work, the selected Si diodes were studied under even higher fluences. A SiC JBS diode was characterised for comparison and also because no irradiation of SiC JBS diodes with high-energy protons has so far been reported. %, despite the potential of these devices for use in radiation environment, irradiation reports are scarce (to the best of our knowledge, only 6 works exist \cite{luo2003impact,witulski2018single,kozlovski2017impact,mochizuki2009influence,vobecky2015impact,kozlovski2018electrical}, among which only one \cite{kozlovski2018electrical} discusses the proton irradiation of high-power SiC JBS diodes).

\section{Components under test}
\label{sec:components}
\subsection{Optical components}
%Here list of optical components, provide Goodfellow references, and sample dimension. Say the number of sample per fluence. Probably a table is the best thing.
Samples of borosilicate (BS), fused quartz (FQ), UV-grade polymethylmethacrylate (PMMA), and polycarbonate (PC) have been tested against $\gamma$-ray irradiation. % We chose to study %the behaviour upon irradiation of 
%these materials as they are commonly adopted in commercial luminaires: glass (BS or FQ) for protective windows, and plastic (PMMA or PC) for secondary optics. 
The samples have been provided by Goodfellow Cambridge Ltd. (UK); manufacturer part references are: LS475409/7, LS475409/3, LS475409/12, LS475409/13 for BS, FQ, PMMA, and PC samples, respectively. All samples were in form of polished disks with diameter of 40~mm and thickness of 3~mm (except BS and PMMA samples, having thickness of 3.3 and 3.1~mm respectively). All materials under test are commercial-grade and not specifically produced for use in radiation environment.

\subsection{Diodes}
Silicon bridge rectifiers and SiC Junction Barrier Schottky (JBS) diodes have been characterized against displacement damage using 24~GeV/\emph{c} protons, to qualify their use in rad-hard AC/DC power supplies for LED lights. The Si diode bridge under test is a Vishay B380C1000504H glass-passivated single-phase bridge rectifier in a WOG package, rated 1~A in average forward current and 600~V in maximum repetitive peak reverse voltage. It shares the same technology %and process methods 
as the Vishay GBU8K diode bridge \cite{b380c2013datasheet,gbu8k2015datasheet}, which has been previously tested against 1~MeV neq fluences up to 8$\times$10$^\text{13}$~n/cm$^\text{2}$ \cite{bager2002lhcb}; for this reason, it was selected as a candidate component. 
Two types of 4H-SiC JBS diodes have also been tested: STPSC10H065D and STPSC10H12-Y diodes from STMicroelectronics, rated 650~V and 1.2~kV in maximum repetitive peak reverse voltage, respectively;  they are both rated for 10~A average forward current. More information on their device structures is available in \cite{st2013sic}. The TO-220 package variants were used to facilitate testing. %Here list of diodes tested, their feature, size of package and technology.

\section{Irradiation conditions}
\label{sec:irradiation}
\subsection{$\gamma$-ray irradiation}
$\gamma$-ray irradiation of optical components was performed at Ionisos (Dagneux, France) \cite{rouif2005radiation}, using a $^\text{60}$Co source. Irradiation was done in air, at room temperature and atmospheric pressure. Four different target doses were specified: 2, 25, 50, and 100~kGy; the actual values reached (as measured by radiochromic dosimeters) are respectively 2.5, 28.5, 58.2, and 113.1~kGy. For the sake of brevity, we will make reference to the nominal values in the remainder of this paper. The doses are specified with respect to water (i.e. in kGy[H$_2$O]).
Table \ref{tab:optical} illustrates the number of samples tested at each dose.
%All samples irradiated to the same dose were contained in a stainless steel perforated basket.
During irradiation, the samples  were contained in stainless steel perforated baskets.
Irradiation of samples with target dose of 2~kGy was performed through 5 complete turns on a secondary tray conveyor around the $^\text{60}$Co source, at a dose rate comprised between 0.5--1~kGy/h. %were loaded on a secondary conveyor belt passing close to the $^{60}$Co source; the target dose of 2~kGy was reached after 5 passages, at a dose rate comprised between 0.5--1~kGy. 
Samples with 25, 50, and 100~kGy target doses were irradiated statically (i.e. placed on metallic shelves facing the source), at average dose rates of 0.46~kGy/h, 0.6~kGy/h, and 0.58~kGy/h, respectively. Irradiation of samples with the highest target doses started first, so that irradiation at all doses finished simultaneously, avoiding different annealing times for different target doses. % Irradiation at different doses finished at the same time, by starting irradiation of samples with highest target doses first; in this way, different annealing times for different doses were avoided. %All target doses were reached at the same time.   
% Irradiation of  doses, dose rates, annealing at ambient temperature, test finished at the same time to reduce the impact of annealing phenomena. 
The irradiated samples were stored at room temperature, and were all characterized within 50 to 52 hours from completion of irradiation. 
%Dire magari che erano in box metallici arieggiati. e che c'erano tanti dosimetri in ogni box.

\begin{table}[!hb]
	\centering
	\caption{Number of samples irradiated per dose}
	\label{tab:optical}
	\begin{tabular}{ccccc} \toprule
		%		\multirow{2}{*}{Material} & \multicolumn{4}{c}{Number of samples irradiated per dose} \\
		%		 & 2 kGy & 25 kGy & 50 kGy & 100 kGy \\ \midrule
		Material  & 2 kGy & 25 kGy & 50 kGy & 100 kGy \\ \midrule
		BS & 1 pc. & 1 pc. & -- & 1 pc. \\ 
		FQ & 1 pc. & 1 pc. & -- & 1 pc. \\
		PMMA & 4 pcs. & 4 pcs. & 4 pcs. & 4 pcs. \\
		PC  & 4 pcs. & 4 pcs. & 4 pcs. & 4 pcs. \\ \bottomrule
	\end{tabular}
\end{table}

\subsection{24~GeV/\emph{c} proton irradiation}
Irradiation of samples of Si and SiC diodes was done at IRRAD facility (CERN) \cite{ravotti2015irrad}, using a 24~GeV/\emph{c} proton beam having a nominal gaussian profile with a FWHM of 1.2$\times$1.2~cm$^\text{2}$. The actual proton fluence was determined by measuring the activation of thin aluminium foils placed in front of the samples during irradiation. A fluence of 24~GeV/\emph{c} protons can be converted into a 1~MeV neq fluence providing the same non-ionising energy loss in silicon through the factor of 0.58 \cite{isidre2016niel}.
%Each individual diode was fixed with Kapton tape to a cardboard holder; the cardboards were mounted on a sample holder which was placed into the irradiating beam through a remotely-controlled conveyor (\textquotedblleft shuttle"), called IRRAD1. 
Irradiation was done at room temperature and atmospheric pressure. The samples were not powered while irradiated (passive test). Two batches of both Si and SiC diodes were irradiated; the target and actual fluences are collected in Table \ref{tab:semi}. Each batch comprises 5 diodes, except batch \#1 of Si diodes, which included only 3 units. After irradiation, the diodes were stored at ambient temperature. 
%For Si diode bridge, two nominal fluences: 8e12 (9.5e12 - 20 min irrad time; annealing 6 days); 3e14 (2.4e14 - 9h e 1/2 irrad time; annealing 5 days). For SiC Schottky diodes, two nominal fluences: 1e13 (1.2e13 - 20 mins - annealing 20 days); 2e14 (1.9e14 - 4 hours; 38 days anneal time).  
Measurements of irradiated diodes were taken only upon authorisation of the irradiation facility personnel, when the residual activation of the samples was acceptably low. 
The time before measurement (i.e. annealing time) for batches \#1 and \#2 of Si diodes was 6 and 5 days, respectively; for SiC diodes, it was 20 and 38 days, respectively. % Dire IRRAD1 shuttle. Dire quanto e' durata l'irradiazione. Dire il tempo di annealing. Dire tutto a temperatura ambiente. Dire che sono test passivi. Dire del beam gaussian profile 1.2x1.2 cm FWHM.
Radiation damage in 4H-SiC is known to be thermally stable at room temperature \cite{lebedev2000doping} and should not be affected by the longer annealing period, although some form of recovery (associated to unstable radiation-induced defects) may not be excluded \cite{doyle1998electrically}.

\begin{table}[!h]
	\centering
	\renewcommand{\arraystretch}{1.0}
	\caption{Irradiation fluences of Si and SiC diodes}
	\label{tab:semi}
	\begin{tabular}{cccc} \toprule
		\begin{tabular}{@{}c@{}}Device\\under test\end{tabular} & \begin{tabular}{@{}c@{}} Target 24 GeV/\emph{c}\\ fluence\end{tabular} & \begin{tabular}{@{}c@{}} Actual  24 GeV/\emph{c} \\ fluence\end{tabular} & \begin{tabular}{@{}c@{}} Actual 1 MeV \\neq fluence in Si\end{tabular} \\ \midrule
		\begin{tabular}{@{}c@{}}Si diodes\\(batch \#1) \end{tabular} & 8$\times$10$^\text{12}$ p/cm$^\text{2}$ & 9.5$\times$10$^\text{12}$ p/cm$^\text{2}$ & 5.5$\times$10$^\text{12}$ n/cm$^\text{2}$ \\ [0.25cm]
		\begin{tabular}{@{}c@{}}Si diodes\\(batch \#2)\end{tabular} & 3$\times$10$^\text{14}$ p/cm$^\text{2}$ & 2.4$\times$10$^\text{14}$ p/cm$^\text{2}$ & 1.4$\times$10$^\text{14}$ n/cm$^\text{2}$ \\ [0.25cm]
		\begin{tabular}{@{}c@{}}SiC diodes\\(batch \#1)\end{tabular} & 10$^\text{13}$ p/cm$^\text{2}$ & 1.2$\times$10$^\text{13}$ p/cm$^\text{2}$ & 7$\times$10$^\text{12}$ n/cm$^\text{2}$ \\ [0.25cm]
		\begin{tabular}{@{}c@{}}SiC diodes\\(batch \#2)\end{tabular} & 2$\times$10$^\text{14}$ p/cm$^\text{2}$ & 1.9$\times$10$^\text{14}$ p/cm$^\text{2}$ & 1.1$\times$10$^\text{14}$ n/cm$^\text{2}$ \\ \bottomrule
	\end{tabular}
\end{table}

\section{Experimental results}
\label{sec:results}
\subsection{Optical components}
\begin{figure}[!t]
	\centering
	\subfloat[]{\includegraphics[scale=0.4]{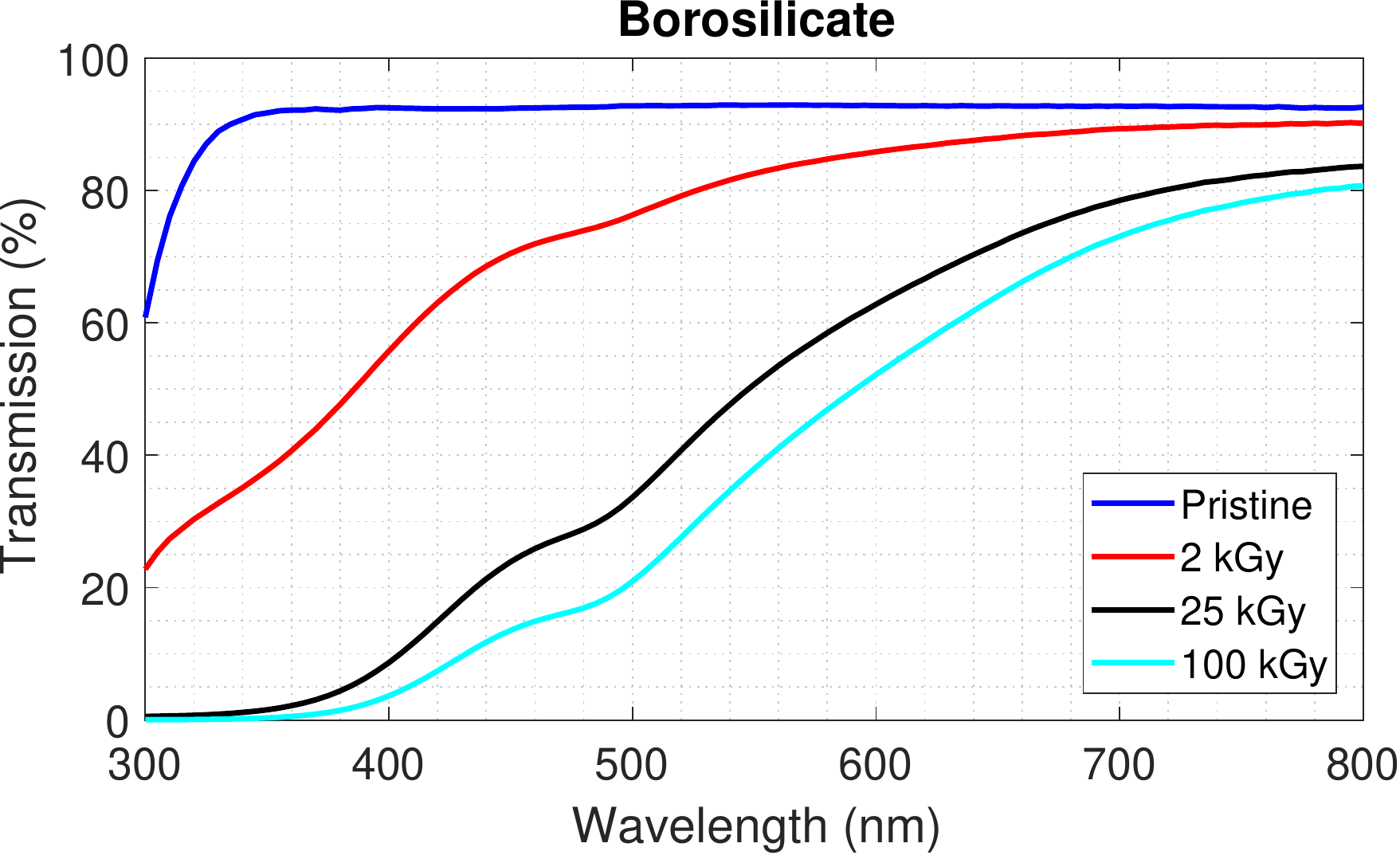}\label{fig:transm:boro}}\hfill
	\subfloat[]{\includegraphics[scale=0.4]{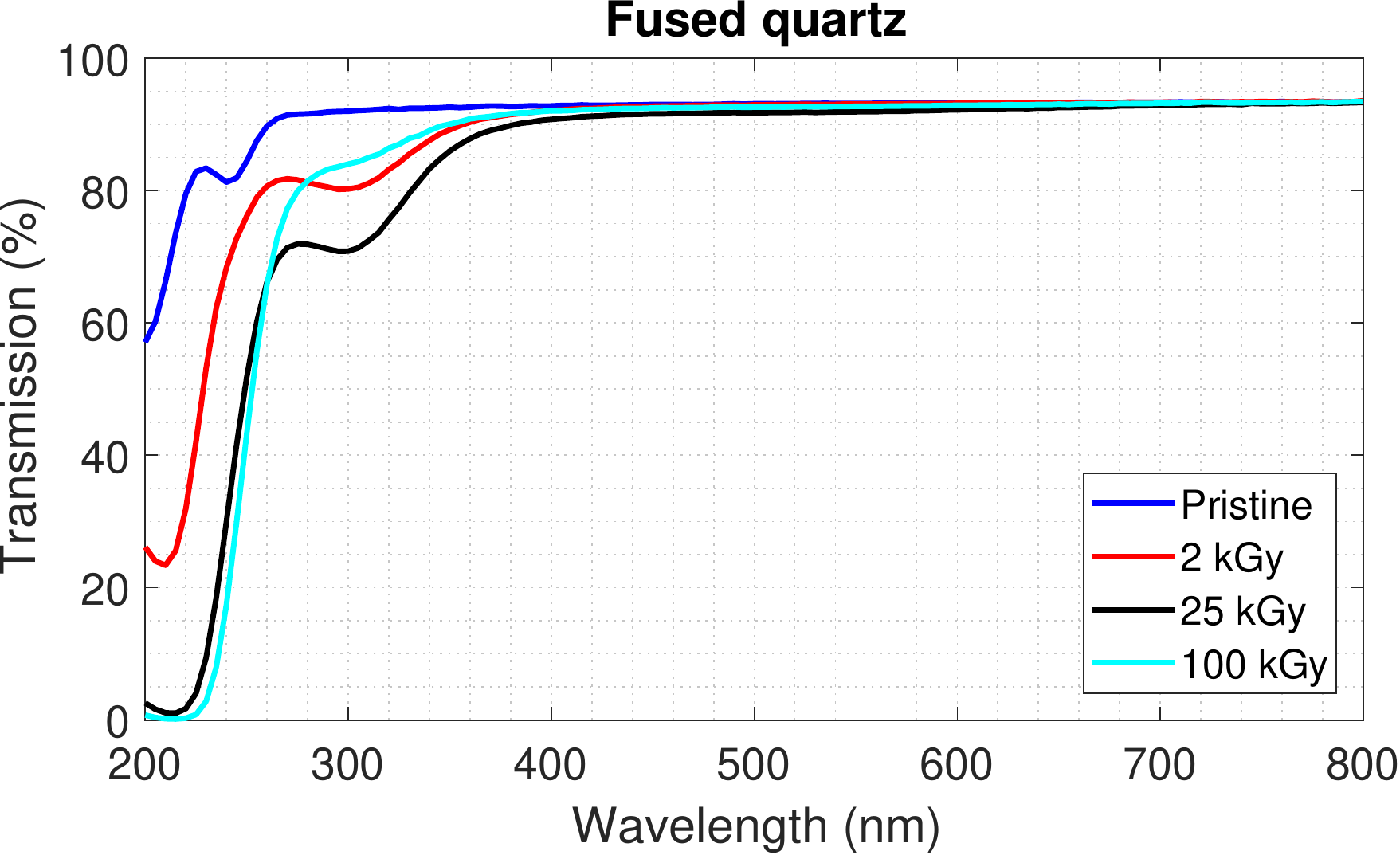}\label{fig:transm:quartz}}\hfill
	\subfloat[]{\includegraphics[scale=0.4]{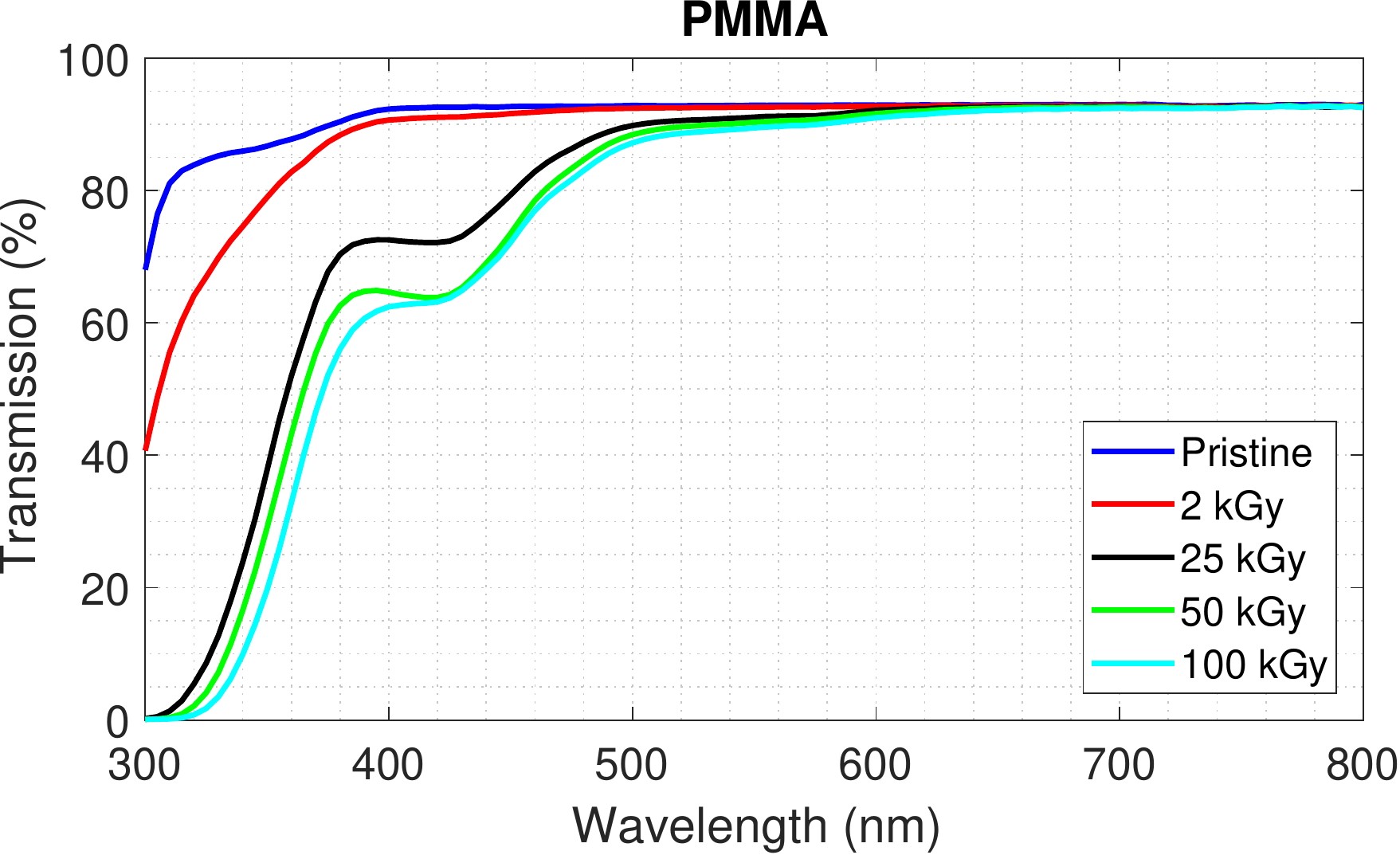}\label{fig:transm:PMMA}}\hfill
	\subfloat[]{\includegraphics[scale=0.4]{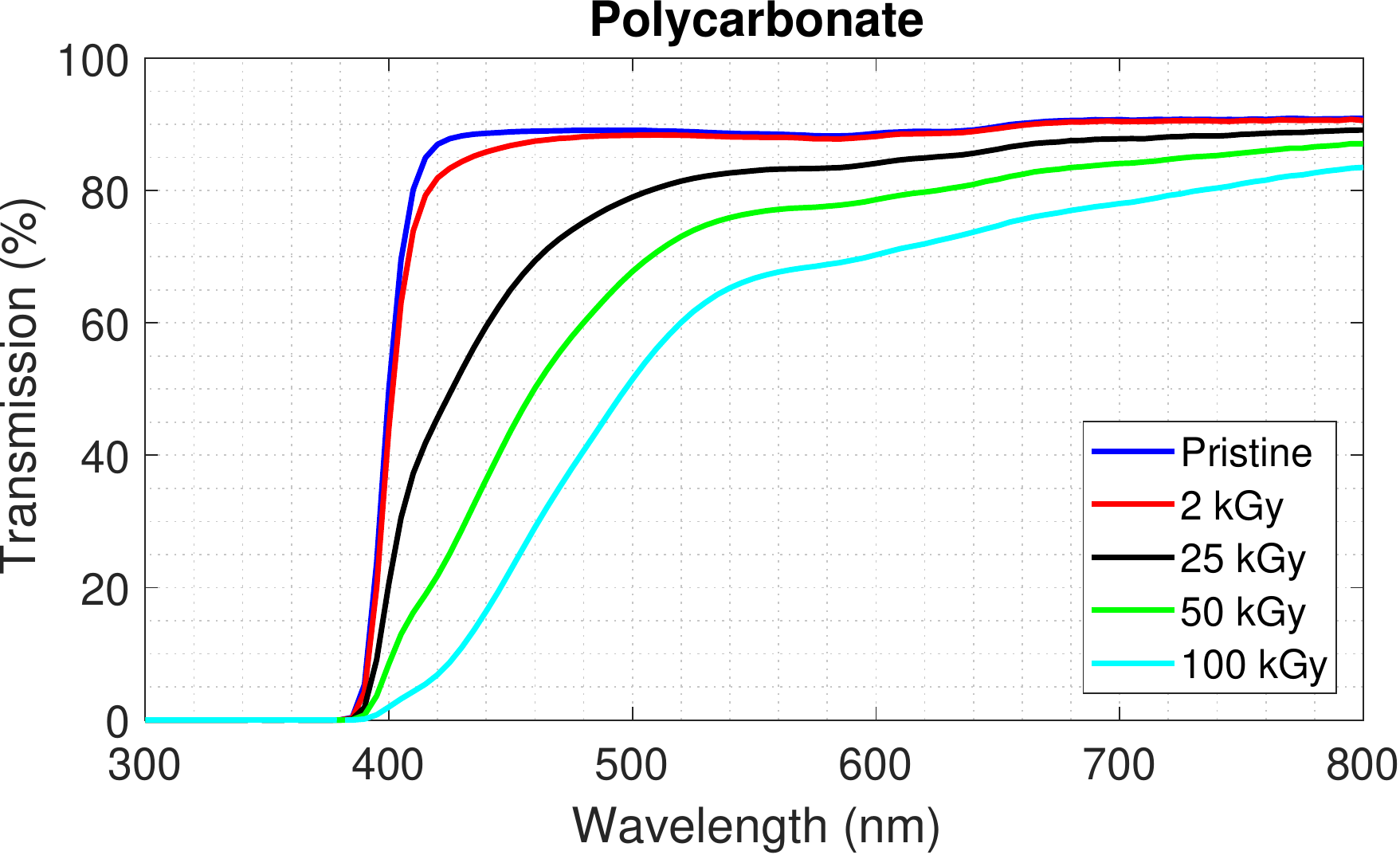}\label{fig:transm:PC}}\hfill
	\caption{Transmission spectra before and after $\gamma$-ray irradiation of samples of: (a) borosilicate, (b) fused quartz, (c) PMMA, (d) polycarbonate.}
	\label{fig:transm}
\end{figure}

Transmission spectra of all samples of optical material were recorded before and after irradiation using a Perkin-Elmer UV-VIS Lambda 650 spectrophotometer equipped with a 150~mm integrating sphere in the range 300--800~nm (only for quartz samples: 200--800~nm). Figure \ref{fig:transm} illustrates the transmission spectra before and after $\gamma$-ray exposure. Equivalent results were obtained for all PMMA and PC samples irradiated to the same dose, so data relative to only one sample per dose is presented for these materials.

\begin{figure}
	\centering
	\subfloat[]{\includegraphics[width=0.8\columnwidth]{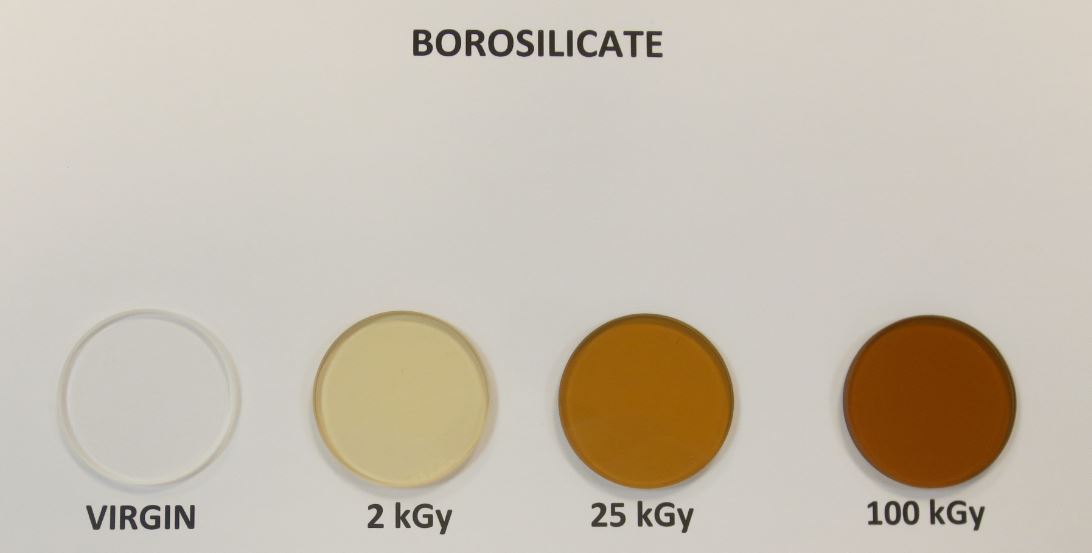}\label{fig:pic:boro}}\hfill
	\subfloat[]{\includegraphics[width=0.8\columnwidth]{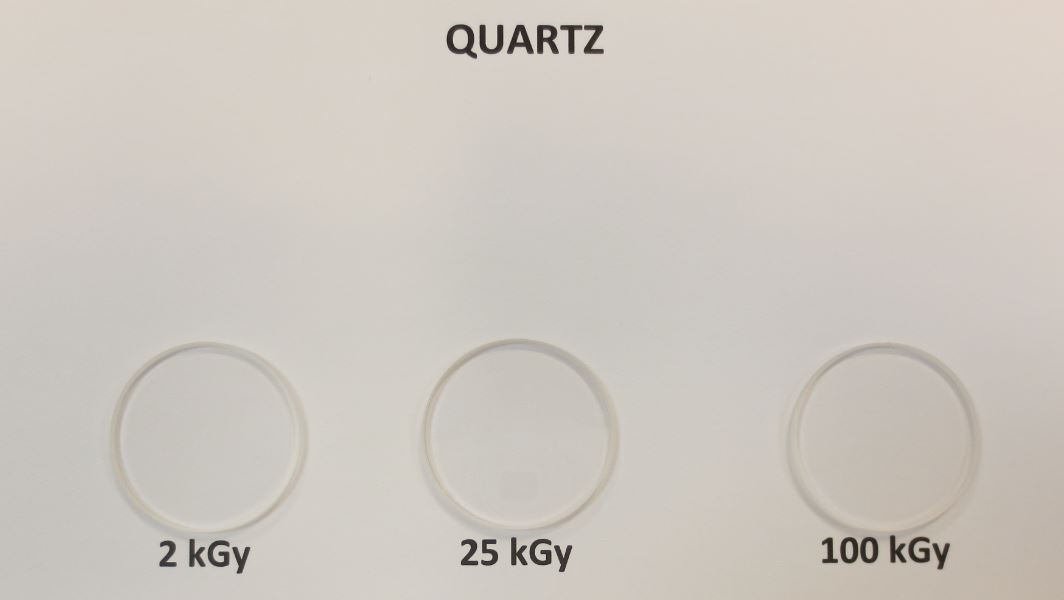}\label{fig:pic:quartz}}\hfill
	\subfloat[]{\includegraphics[width=0.8\columnwidth]{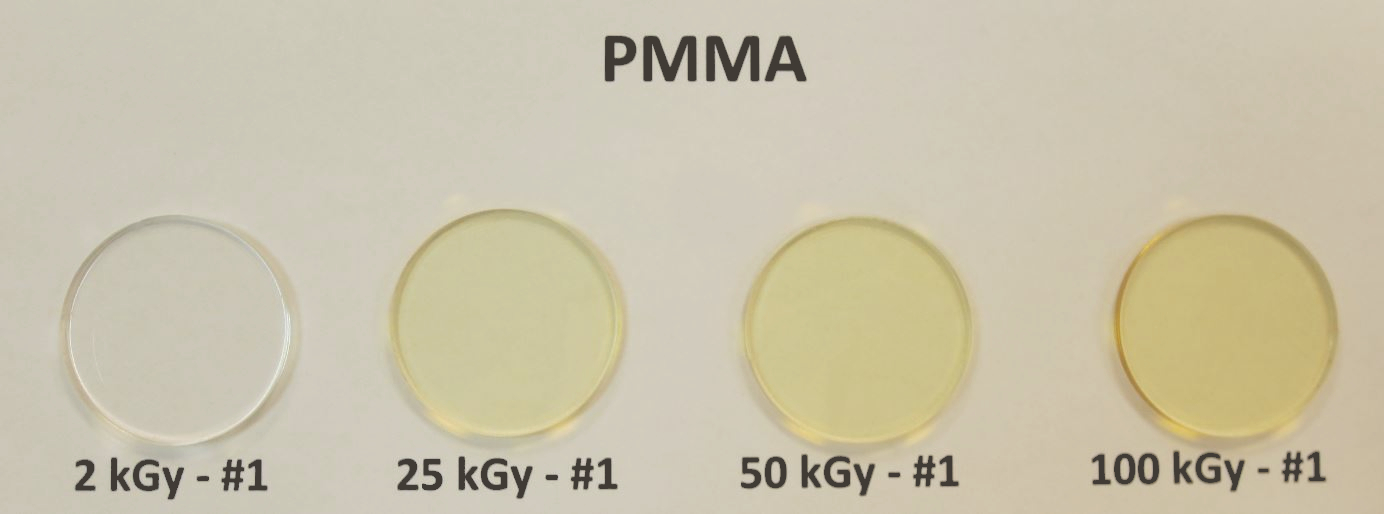}\label{fig:pic:PMMA}}\hfill
	\subfloat[]{\includegraphics[width=0.8\columnwidth]{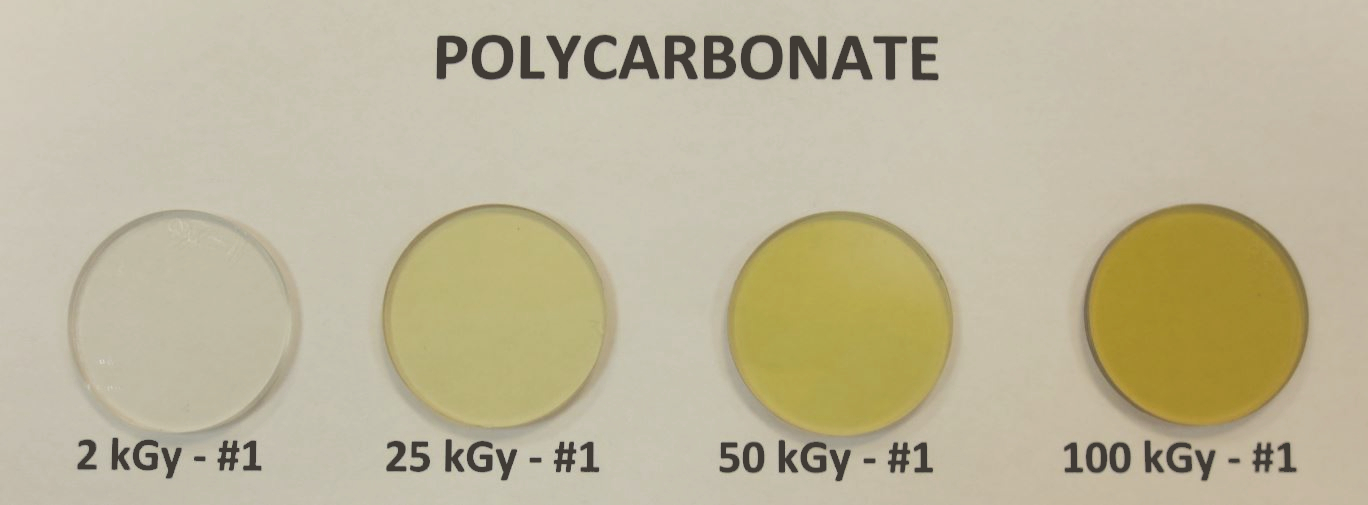}\label{fig:pic:PC}}\hfill
	\caption{Pictures of irradiated samples of: (a) borosilicate, (b) fused quartz, (c) PMMA, (d) polycarbonate.}
	\label{fig:pictures}
\end{figure}

Before irradiation, transmittance of BS is $>$90\% (see Fig. \ref{fig:transm:boro}) until the UV cut-off edge at 335~nm \cite{wang2012radiation,wang2009effects}. Upon increasing dose, two strong absorption bands located near 330~nm and 475~nm appear; these bands have been previously attributed to boron oxygen hole centres \cite{wang2016radiation} (each centre is given by a hole trapped on a bridge oxygen structure with [BO$_4$]$^-$). The large increase of absorption in the UV-blue portion of the spectrum leads to the darkening of BS samples (see Fig. \ref{fig:pic:boro}).  

All samples of FQ before irradiation exhibit high transmittance ($>$90\%) until 260~nm (see Fig. \ref{fig:transm:quartz}). Before irradiation, an absorption band located at 240~nm is clearly apparent, which we identify with the $B_2\beta$ band due to Ge oxygen deficient centres (GeODC) \cite{marshall1997induced,leon2009gamma}; this band precedes the UV cut-off edge, occurring at values close to those reported in literature for other samples of FQ \cite{marshall1997induced,leon2009gamma}. After irradiation, two new absorption bands emerge in the UV: the first at $\sim$210~nm, the second at 300~nm. The first band is attributed to the E$^\prime$ centre (Si dangling bond $\equiv$Si$\cdot$) \cite{marshall1997induced,leon2009gamma};  with increasing dose, absorption at E$^\prime$ centres increases steadily, and slightly shifts to longer wavelengths. The origin of the absorption band at 300~nm, referred to in literature as the $B_1$ band, is still unclear \cite{marshall1997induced,leon2009gamma}, although it is thought to be formed from the conversion of the $B_2\beta$ band: the proposed mechanism is that a GeODC loses an electron due to irradiation creating a centre absorbing at 300~nm \cite{marshall1997induced}. 
It has been demonstrated in previous works \cite{marshall1997induced,guzzi1993neutron} that the emergence of the $B_1$ band  upon irradiation occurs only in FQ, and not in crystalline quartz or synthetic fused silica.
Absorption at this band increases with increasing dose up to 25~kGy; however, it decreases after 100~kGy. 
In previous works on the irradiation of FQ using neutrons and x-rays \cite{guzzi1993neutron,mitchell1956cxi,paige1957kinetics,levy1955radiation}, it was shown that, in some specimens of FQ, the $B_1$ band initially increased with increasing radiation exposure, but after reaching a maximum, it started to decrease with further irradiation (this phenomenon is termed radiation bleaching). A detailed analysis of this phenomenon was performed in \cite{paige1957kinetics}, where radiation bleaching of the $B_1$ band was explained in terms of interdependence between the occupancy of $B_1$ centres (i.e. the traps responsible for the $B_1$ band) and the occupancy of the other defects present in irradiated FQ. We therefore attribute the decrease in absorption of the $B_1$ band observed in our sample irradiated at 100~kGy to the same phenomenon already observed in the case of neutron and x-rays irradiation.
%This fact can be due to a different concentration of precursors (intrinsic defects and impurities) in the sample irradiated at 100~kGy, or to the rise of other absorption mechanisms distinct from the conversion of the $B_2\beta$ band into the 300~nm band. 
In any case, the FQ samples stay transparent in the visible range even after the highest dose, as can be visually seen from Fig. \ref{fig:pic:quartz}. 

The transmittance curves of $\gamma$-irradiated UV-grade PMMA samples are shown in Fig. \ref{fig:transm:PMMA}. Upon irradiation, a strong increase in UV absorption occurs, and two absorption bands between 385--425~nm and 515--580~nm emerge. Increased absorption in the UV has been attributed to increased density of dienes group (--C$=$C--)$_n$ due to radiation \cite{rai2017uv}. The two bands in the visible are due to radiation-induced free radicals acting as colour centres \cite{rai2017uv,clough1996color,lu2000transmittance}; $\gamma$-ray irradiation involves scission of C$=$O, C--C, C--O--C, and C--C--O bonds \cite{rai2017uv}. It is also evident that radiation-induced optical damage tends to saturate at higher doses (the curves after 50 and 100~kGy are almost overlapping); this phenomenon is due to competition between molecular scission and cross-linking induced by radiation \cite{rai2017uv}. Previous studies in literature have shown that cross-linking under $\gamma$-ray irradiation occurs at doses $>$10~kGy, and is initiated by the absorption of water from airborne moisture \cite{rai2017uv}. Cross-linking occurs between C--H, O--H, and H--O--H bonds (the last two induced by absorbed water molecules) \cite{rai2017uv}. 
%A strong increase in the UV cut-off UV: diene C=C; poi gli altri due sono centri di colore dovuti a molecular scission=radical formation; i due colour centres si vedono abbastanza (il primo sì, il secondo nelle ref è piu una banda che gradualmente si raccorda al valore del sample pristine) nella prima e terza ref. Il motivo x cui si satura è xk c'è bilancio tra cross-linking e molecular scissions. Cross linking iniziale dovuto ad assorbimento di acqua. La scissione riguarda principalmente C=O, C-C, C-O-C, C-C-O, con cross-linking di CH, OH, e HOH (gli ultimi due dovuti all'acqua).  %Non si dice però quale radicale (=quale scissione) sia responsabile dei centri di colore.
% molecular scission (free radical formation) 
As a consequence of the aforementioned colour centres, PMMA samples turned yellow upon irradiation (see Fig. \ref{fig:pic:PMMA}).

Finally, the transmittance curves of PC samples are shown in Fig. \ref{fig:transm:PC}. Several absorption bands are introduced by radiation in the visible range, the most evident ones being around $\sim$415~nm and between 550--580~nm. 
Previous works \cite{sharma2008modification,sinha2004gamma,factor1994chemistry,factor1997use} have shown that $\gamma$-ray irradiation of PC leads to chain scissioning and formation of %generation of free radicals which react to form 
different types of products (like e.g. species containing ortho-quinone groups), acting as colour centres in the visible spectrum. Unlike PMMA, absorption at these radiation-induced bands does not saturate at high doses (conversely, the optical damage continues to increase with dose). 
%As in PMMA, these bands are due to radiation-induced free radicals acting as colour centres \cite{clough1996color,sharma2008modification}, however, unlike PMMA, absorption at these bands does not saturate at high doses (conversely, the optical damage continues to increase with dose). 
A larger reduction in trasmittance is obtained in PC than in PMMA, as also demonstrated by the more intense yellow radiation-induced colouration of the PC samples (see Fig. \ref{fig:pic:PC}).  

Radiation-induced absorption bands in the considered materials decrease if subjected to thermal treatment with high temperatures (i.e. radiation-induced defects originating these bands can be thermally annealed); 
	in particular, previous experiments have shown that recovery of transmission at radiation-induced bands occurs even at room temperature. For more details on thermal annealing at room temperature in air of borosilicate, %(?after gamma irradiation?), 
	we defer to \cite{wang2012radiation,gusarov2005comparison}; for quartz, we defer to \cite{agnello2006thermal} and references therein; for PMMA and PC, we make reference to \cite{clough1996color,clough1995discoloration}. 
	In addition, radiation-induced absorption bands in these materials can also be optically bleached: absorption bands at 210~nm and 300~nm in $\gamma$-ray irradiated FQ are known to photobleach under UV light \cite{levy1955radiation}; visible light can bleach radiation-induced absorption bands in both BS \cite{treadaway1975transient,treadaway1976radiation} and PC \cite{lundy34methods,hermanson1993physical}; optical bleaching of PMMA presents a more complex behaviour, exhibiting a reduction or an enhancement of the radiation-induced absorption in the UV spectrum depending on the emission wavelength of the light source used for bleaching  \cite{chadwick1972effect}.   
	Therefore, keeping in mind the application of these materials as optical components for rad-hard lighting systems, a recovery of the transmission properties is to be expected after irradiation as a consequence of thermal annealing (even at room temperature) and, particularly in BS and PC, optical bleaching induced by the white LEDs (for reference, the electroluminescence spectrum of a white GaN LED extends from 400~nm to 750~nm). Nevertheless, in the design of a rad-hard luminaire, the optical components should be chosen based on their transmission properties just after irradiation (i.e. based on the data presented in the paragraphs above), as this would represent the worst-case (most-conservative) scenario. 
	For example, access by human operators into CERN experimental facilities tunnels is prohibited while the particle accelerators are running, owing to the extremely high radiation levels produced by the circulating beams \cite{de2015radiation}; under such circumstances, the tunnel illumination system is therefore not required to be activated (i.e. lights are off while the accelerators are running). 
	Access to accelerator tunnels is possible only during machine stops, when there is no circulating beam, after a cool down period to allow the residual radiation to decay (typically a few hours from the beam stop, but it can be 30--40~hours for the most radioactive areas). % by human operators, and hence need for lighting, to only during technical stops. In such situation, 
	Consequently, the lighting system shall be designed to provide the stipulated minimum illuminance levels just after the accelerator stops, so that the optical materials offering the best performance just after irradiation should be selected for use in rad-hard luminaires.

In conclusion, BS exhibits a significant reduction of visible light transmission even after 2~kGy (corresponding to $\sim$2 years on the tunnel walls of a typical section of CERN accelerator complex), while FQ retains transparency to visible light even after 100~kGy; therefore, BS protective windows in luminaires for radioactive environment should be avoided in favour to FQ components. Visible light transmission in both PMMA and PC is degraded by radiation-induced colour centres, but larger values of absorption are measured in PC samples; consequently PMMA should be specified instead of PC as a component for secondary optics lenses.

\subsection{Diodes}
%Si diodes: \cite{davies2006radiation,messenger2003niel,spieler1997introduction,bager2002lhcb}. SiC Schottky diodes \cite{harris2005displacement,nava2008silicon,messenger2003niel,lee2003comparative}.
Current-voltage (\emph{I}--\emph{V}) characteristics of Si and SiC diodes were measured before and after irradiation with a Keithley 2410 source-meter unit, at room temperature. 
Measurements of the forward and reverse \emph{I}--\emph{V} curves were done in the range 0--1~A, and $-$20--0~V, respectively. In the case of Si bridge rectifiers, all four diodes contained in each package were tested. Equivalent results were obtained for all diodes irradiated in the same batch; for this reason, data of only one representative sample per fluence is shown. All fluences are given as 1~MeV neq fluences in Si.
% ovviamente intendo per i B380C che ho stessi risultati per tutti i diodi di tutti i package testati alla stessa fluenza.

\begin{figure}
	\centering
	\subfloat[]{\includegraphics[scale=0.45]{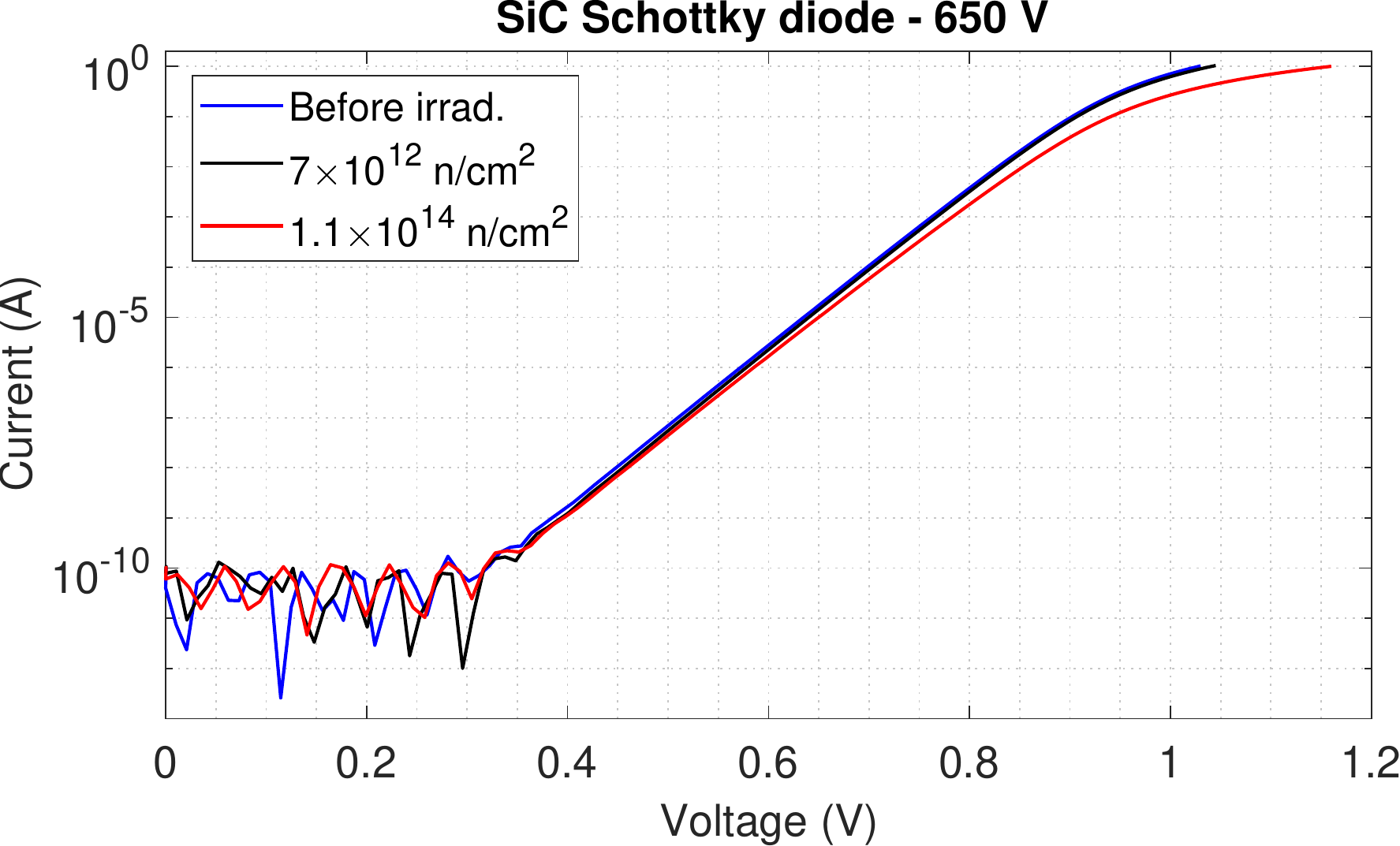}\label{fig:SiC650}}\hfill
	\subfloat[]{\includegraphics[scale=0.45]{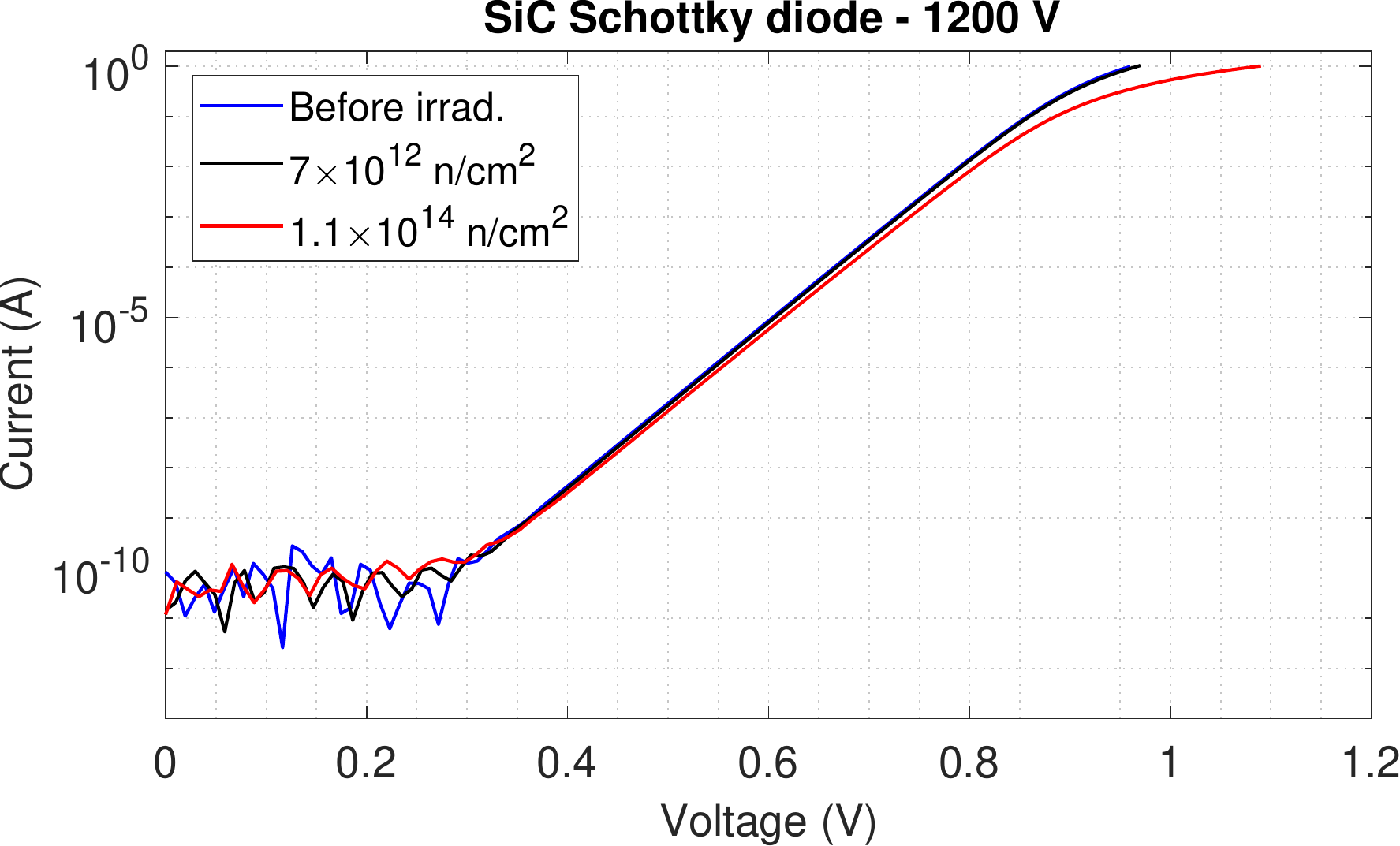}\label{fig:SiC1200}}
	\caption{\emph{I}--\emph{V} characteristics at room temperature of (a) STPSC10H065D and (b) STPSC10H12-Y SiC diodes, before and after irradiation (expressed as 1~MeV neq fluence).}
	\label{fig:SiC}
\end{figure}

\begin{figure}
	\centering
	\subfloat[]{\includegraphics[scale=0.45]{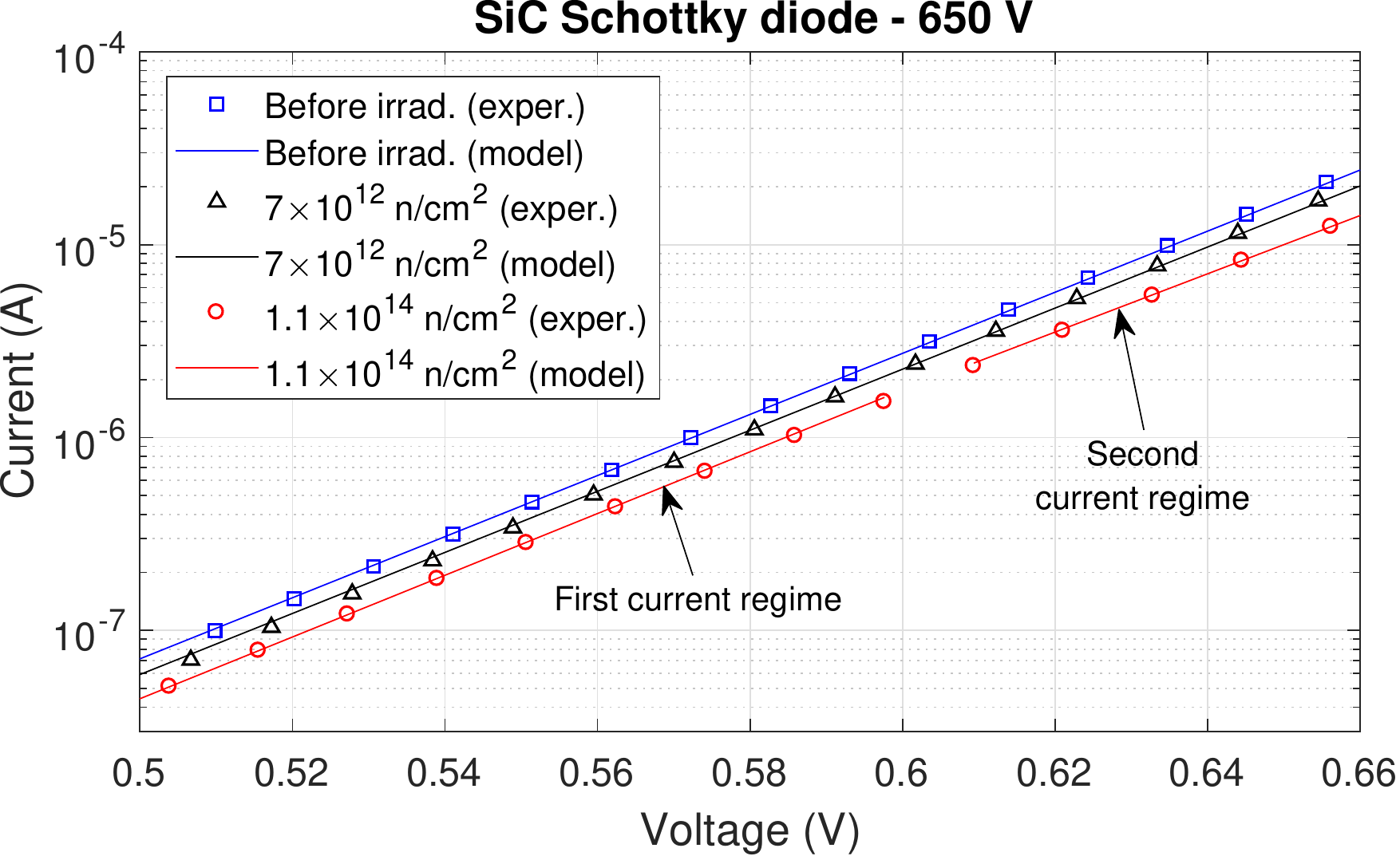}\label{fig:SiC650model}}\hfill
	\subfloat[]{\includegraphics[scale=0.45]{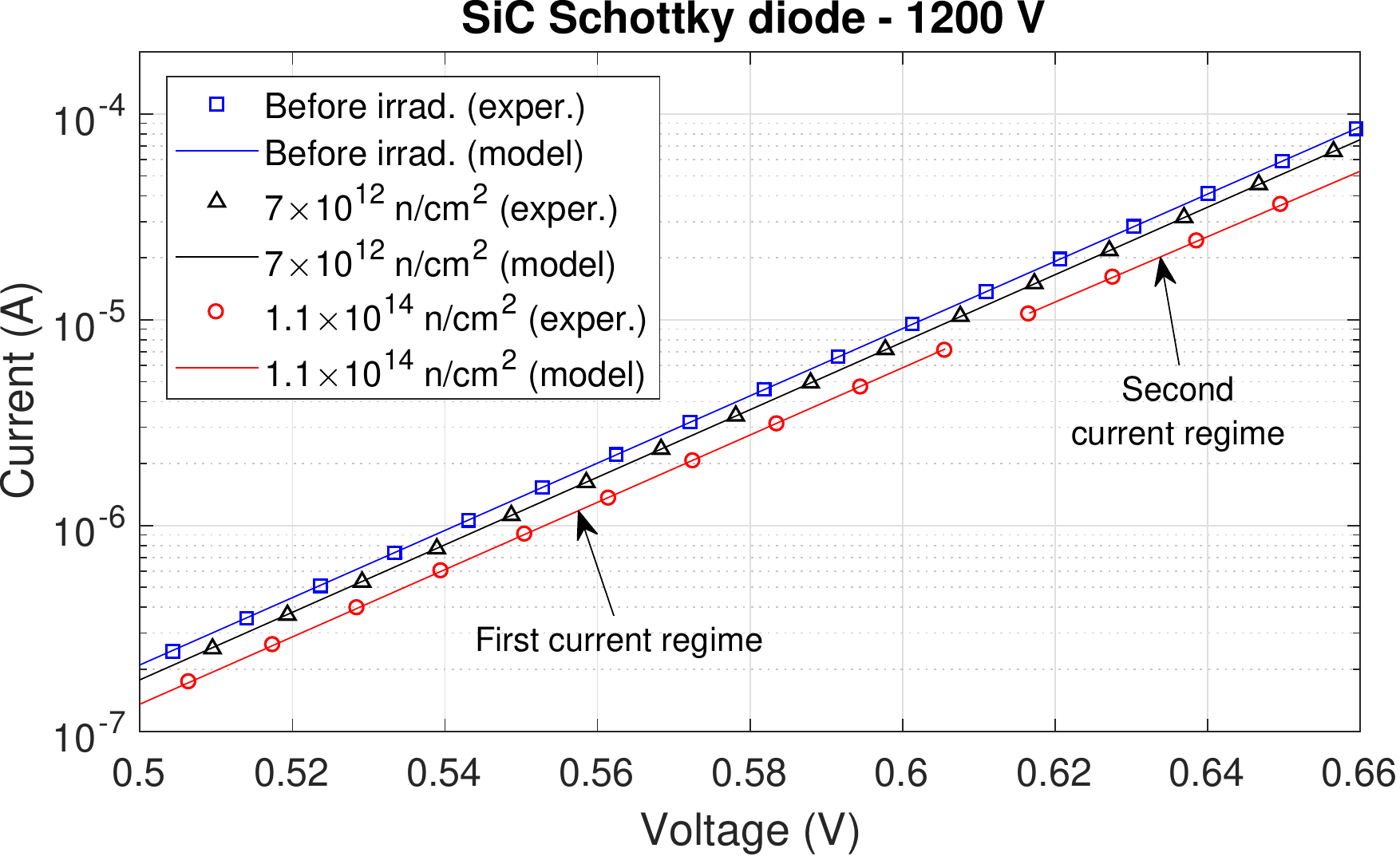}\label{fig:SiC1200model}}
	\caption{Experimental and fitted \emph{I}--\emph{V} characteristics at room temperature of (a) STPSC10H065D and (b) STPSC10H12-Y SiC diodes, before and after irradiation (expressed as 1~MeV neq fluence).}
	\label{fig:SiC:model}
\end{figure}

Figures \ref{fig:SiC650} and \ref{fig:SiC1200} show the forward \emph{I}--\emph{V} curves before and after irradiation of 650~V and 1200~V SiC diodes, respectively. The behaviour of irradiated 650~V and 1200~V SiC diodes was found to be similar: degradation after 7$\times$10$^\text{12}$~n/cm$^\text{2}$ is very small, while larger damage is detected after the highest fluence (in particular, for currents above 0.1~A).
We found that \emph{I}--\emph{V} curves of both 650~V and 1200~V SiC diodes before irradiation and after a fluence of 7$\times$10$^\text{12}$~n/cm$^\text{2}$ can be well described by the standard model of an ideal diode with a series resistance \cite{harris2005displacement}:
\begin{equation}
I=I_s \left\lbrace \exp \left[ \frac{q(V - R_S I)}{\eta k_B T} \right] -1 \right\rbrace, 
\label{eq:IV1}
\end{equation}
where $I_s$ is the saturation current, $q$ is the absolute electronic charge, $\eta$ is the ideality factor, $k_B$ is the Boltzmann constant, $T$ is the absolute temperature, and $R_S$ denotes the series resistance.
On the other hand, after the highest fluence, two segments with different slopes are observed in the \emph{I}--\emph{V} characteristics instead of a single exponential behaviour as in \eqref{eq:IV1}; these two segments represent two different current regimes (i.e. different conduction mechanisms), each characterised by a distinct ideality factor $\eta$. % curve found in the \emph{I}--\emph{V} curve before irradiation or after 7$\times$10$^{12}$~n/cm$^\text{2}$. 
In this case, the \emph{I}--\emph{V} curves of SiC diodes irradiated at 1.1$\times$10$^\text{14}$~n/cm$^\text{2}$ can be expressed as:
\begin{equation}
I=\begin{cases}
I_{s_1}\left[ \exp\left(qV'/\eta_1 k_B T \right) -1 \right], \quad V<V_T
\\ 
I_{s_2} \left[ \exp \left( qV'/\eta_2 k_B T \right) -1 \right], \quad V>V_T
\end{cases}
\label{eq:IV2}
\end{equation}
where $V'=V-R_SI$, and $V_T$ denotes the voltage at which the transition between the two current regimes occurs. % (/the transition voltage between the two current regimes)
For all 650~V and 1200~V SiC diodes tested, $V_T$ is about 0.6~V.

The experimental \emph{I}--\emph{V} characteristics of SiC diodes were fitted to their corresponding models \eqref{eq:IV1} or \eqref{eq:IV2}. Before irradiation and after 7$\times$10$^\text{12}$~n/cm$^\text{2}$, the values of $\eta$ and $I_s$ in \eqref{eq:IV1} have been extracted through an exponential fit in the current range from 5$\times$10$^{-\text{8}}$~A to 5$\times$10$^{-\text{4}}$~A. After 1.1$\times$10$^\text{14}$~n/cm$^\text{2}$, 
	the saturation current and ideality factor for the first (respectively, second) current regime in \eqref{eq:IV2} are obtained by exponential fitting of experimental \emph{I}--\emph{V} curves in the current interval from 10$^{-\text{9}}$~A to 10$^{-\text{6}}$~A (respectively, from 5$\times$10$^{-\text{6}}$~A to 5$\times$10$^{-\text{4}}$~A). 
Analogously to \cite{harris2005displacement}, $R_S$ in \eqref{eq:IV1} and \eqref{eq:IV2} was disregarded while performing these exponential fits, since its effect is negligible in the considered current intervals.
%	 the saturation current and ideality factor for the first and second current regime in \eqref{eq:IV2} are obtained by exponential fitting of experimental \emph{I}--\emph{V} curves in the current intervals from $\sim$10$^\text{-9}$~A to 10$^\text{-6}$~A, and from $\sim$5$\times$10$^\text{-6}$~A to $\sim$5$\times$10$^\text{-4}$~A, respectively. 
$R_S$ was then extracted for all curves by fitting the experimental \emph{I}--\emph{V} characteristics for $I>$0.5~A. 
Figures \ref{fig:SiC650model} and \ref{fig:SiC1200model} illustrate the experimental and fitted \emph{I}--\emph{V} curves for 650~V and 1200~V SiC diodes, respectively; for clarity reasons (i.e. to allow for a distinct representation of all curves), the data is shown only in a limited interval (0.5--0.66~V), which nevertheless spans across almost three orders of magnitude in current. For \emph{I}--\emph{V} curves after 1.1$\times$10$^\text{14}$~n/cm$^\text{2}$, the first and second current regimes are also indicated. As can be seen, fitted and experimental data are in excellent agreement in the represented interval. % (in fact, such agreement extends further, from $\sim$10$^\text{-8}$~A to $\sim$5$\times$10$^\text{-4}$~A).
%Figures \ref{fig:SiC650model}--\ref{fig:SiC1200model} illustrate the experimental and fitted \emph{I}--\emph{V} curves for 650~V and 1200~V SiC diodes, respectively, in the voltage range 0.5--0.66~V (such interval is chosen for clarity reasons, to allow for a distinct representation of all curves in a wide range of currents). As can be seen, fitted and experimental data are in excellent agreement over more than 3 orders of magnitude in current (such agreement extends further). %As also certified by the statistical factors. 

\begin{table}[!b]
	\centering
	\renewcommand{\arraystretch}{1.0}
	\caption{Relative variation of $I_s$, $\eta$, and $R_S$ in SiC diodes upon high-energy proton irradiation}
	\label{tab:SiC}
	\begin{tabular}{ccccc} \toprule
		\begin{tabular}{@{}c@{}}Voltage\\rating\end{tabular} & \begin{tabular}{@{}c@{}}1 MeV neq \\ fluence in Si\end{tabular} & \begin{tabular}{@{}c@{}}$\Delta I_s$ \\ $\left(\Delta I_{s_1}\right)$ \end{tabular} & \begin{tabular}{@{}c@{}}$\Delta \eta$ \\ $\left(\Delta \eta_1\right)$\end{tabular} & $\Delta R_S$  \\ \midrule
		650~V & \begin{tabular}{@{}c@{}}7$\times$10$^\text{12}$ n/cm$^\text{2}$\end{tabular} & $-$16\% & $+$0.03\% &  $+$8\%  \\ [0.25cm]
		650~V & \begin{tabular}{@{}c@{}}1.1$\times$10$^\text{14}$ n/cm$^\text{2}$\end{tabular} & $-$33\% & $+$0.13\% &   $+$105\% \\ [0.25cm]
		1200~V & \begin{tabular}{@{}c@{}}7$\times$10$^\text{12}$ n/cm$^\text{2}$\end{tabular} & $-$21\% & $-$0.38\% &  $+$9\%  \\ [0.25cm]
		1200~V & \begin{tabular}{@{}c@{}}1.1$\times$10$^\text{14}$ n/cm$^\text{2}$\end{tabular} & $-$24\% & $+$0.27\% & $+$143\%  \\ \bottomrule
	\end{tabular}
\end{table}

Radiation introduces in 4H-SiC a wide range of defects, predominantly vacancies and interstitials of Si and C atoms, and combinations thereof (due to low mobility of primary radiation defects, formation of defect complexes including impurities is limited) \cite{nava2008silicon}. These defects act as compensation and trapping centres, leading to a reduction in free electron density (carrier removal) and hence to greater series resistance $R_S$ and lower values of saturation current $I_s$ \cite{luo2003impact,kozlovski2018electrical,nava2008silicon}.
Such variations of $I_s$ and $R_S$ are apparent in Figs. \ref{fig:SiC650} and \ref{fig:SiC1200}, and can be quantitatively analysed by studying the evolution with increasing radiation exposure of the fitted values of %fit variables 
$I_s$, $\eta$ and $R_S$ in models \eqref{eq:IV1} and \eqref{eq:IV2}.
The relative variations of these parameters %the fitted values of $I_s$, $\eta$, and $R_S$ 
upon irradiation %of SiC diodes 
are collected in Table \ref{tab:SiC}; the change in saturation current and ideality factor after the highest fluence is computed with respect to values of the first current regime, i.e. $I_{s_1}$ and $\eta_1$. All variations are computed as $(X_a-X_b)/X_b$, where $X_b$ and $X_a$ denote the same quantity before and after irradiation, respectively.
After a fluence of 7$\times$10$^\text{12}$~n/cm$^\text{2}$, a reduction in $I_s$ of $-$16\% and an increase in $R_S$ of 8\% w.r.t. pre-irradiated samples is found in 650~V SiC diodes; in 1200~V diodes, $I_s$ decreases by $-$21\% and $R_S$ increases by 9\%. In addition, only marginal variations in $\eta$ upon irradiation are found in both types of SiC diodes ($\eta\approx\text{1.074}$ and $\eta\approx\text{1.04}$ for 650~V and 1200~V diodes, respectively, both before and after irradiation). Hence, modest carrier removal is already detected after 7$\times$10$^\text{12}$~n/cm$^\text{2}$, although its impact on the electrical characteristics of the diodes is negligible, as apparent from Figs. \ref{fig:SiC} and \ref{fig:SiC:model}.
After irradiation at 1.1$\times$10$^\text{14}$~n/cm$^\text{2}$, $R_S$ in 650~V diodes increases by $\sim$105\% w.r.t. pre-irradiated samples, changing from 95~m$\Omega$ (before irradiation) to 195~m$\Omega$ (after 1.1$\times$10$^\text{14}$~n/cm$^\text{2}$); $R_S$ in 1200~V diodes degrades even more, increasing by $\sim$143\% from 74~m$\Omega$ (before irradiation) to 180~m$\Omega$ (after 1.1$\times$10$^\text{14}$~n/cm$^\text{2}$). This increase in $R_S$ is responsible for the deviations observed in the \emph{I}--\emph{V} curves at currents $I>$0.1~A after the highest fluence (see Figs. \ref{fig:SiC650} and \ref{fig:SiC1200}). The larger increase of $R_S$ in 1200~V devices is attributed to different device geometry (like e.g. larger vertical thickness of the n-drift region). In any case, this increase in $R_S$ has only a limited effect on the electrical characteristics of SiC diodes for currents up to 1~A. 
%After irradiation at 1.1$\times$10$^{14}$~n/cm$^\text{2}$, $R_S$ in 650~V and 1200~V diodes increases by $\sim$104\% and $\sim$142\% w.r.t. pre-irradiated samples, respectively; this increase in $R_S$ is responsible for the deviations observed in the \emph{I}--\emph{V} curves at currents $I>$0.1~A after the highest fluence (see Figs. \ref{fig:SiC650} and \ref{fig:SiC1200}). The larger increase of $R_S$ in 1200~V devices is attributed to different device geometry (like e.g. larger vertical thickness of the n-drift region). 
The ideality factor of the first region ($\eta_1$) after 1.1$\times$10$^\text{14}$~n/cm$^\text{2}$ is nearly identical to its corresponding pre-irradiation values in both 650~V and 1200~V diodes; in addition, $I_{s_1}$ decreases by $-$33\% and $-$24\% w.r.t. pre-irradiated samples in 650~V and 1200~V diodes, respectively.  
Concerning the second current regime, the values of $\eta_2$ are higher than those of $\eta_1$: $\eta_2=\text{1.126}$ and $\eta_2=\text{1.07}$ in 650~V and 1200~V diodes, respectively, i.e. they are 6.1\% and 3.1\% larger than their corresponding $\eta_1$ values; a slight increase of the ideality factor upon proton irradiation of 4H-SiC JBS diodes was also found in \cite{harris2005displacement}. We attribute the emergence of the second current regime to radiation-induced defects.
In addition, we ascribe the slightly different behaviour of $I_{s_1}$ and $\eta_2$ between 650~V and 1200~V diodes after 1.1$\times$10$^\text{14}$~n/cm$^\text{2}$ to different device geometry and doping characteristics, leading to marginally dissimilar current conduction mechanisms.
%\AF{In addition, after 1.1$\times$10$^{14}$~n/cm$^\text{2}$, two different current regimes (characterised by two distinct ideality factors $\eta$) appear: the first one from $\sim$0.4~V to $\sim$0.5~V, with an ideality factor almost identical to the pre-irradiation one ($\sim$1.074), and the second one from $\sim$0.6~V to $\sim$0.8~V (with a slightly higher $\eta$ of $\sim$1.123). $I_s$ in this first current regime decreases by $-$33\% w.r.t. pre-irradiated samples.} 
% We attribute the emergence of the second current regime to radiation-induced defects. %TO DO: nella versione finale del paper, discutere meglio dell'aumento dell eta; harris 2005 diceva che eta aumenta. Ma forse anche nel report di Glaser e Nava? No; solo Harris.
Our results in terms of device degradation are in good agreement with those in \cite{kozlovski2018electrical,harris2005displacement}, despite use of a diode from a different manufacturer and with different ratings (normalization of radiation damage at various proton energies is done via NIEL curves for SiC obtained using screened Coulomb potential \cite{messenger2003niel,lee2003comparative} and considering the effect of nuclear interactions dominating at energies $>$100~MeV \cite{khanna2004proton}). % With respect to report \cite{kozlovski2018electrical}, radiation damage emerges approximately after the same equivalent fluence; however. 
Finally, no variations are detected in the reverse bias curves, as expected for Schottky-based diodes and as already noted in literature for similar devices tested at comparable fluences \cite{luo2003impact,harris2005displacement}.  %\textbf{sarebbe bello dire a quale fluenza inizia il danno dallo studio sui detector}
All these results show that the 4H-SiC JBS diodes under test present only a limited degradation for currents up to 1~A after fluences of 1.1$\times$10$^\text{14}$~n/cm$^\text{2}$, and hence are expected to reliably withstand the radiation environment of a typical section of CERN accelerator tunnels \cite{de2015radiation} for at least 10 years.

%\AF{TO DO: Dire che l'eta saliva. } % Scrivere i modelli. Dire dell'errore R2.  Dire delle nuove figure. Dire che testavo 5 campioni per fluenza?

\begin{figure}
	\centering
	\subfloat[]{\includegraphics[scale=0.45]{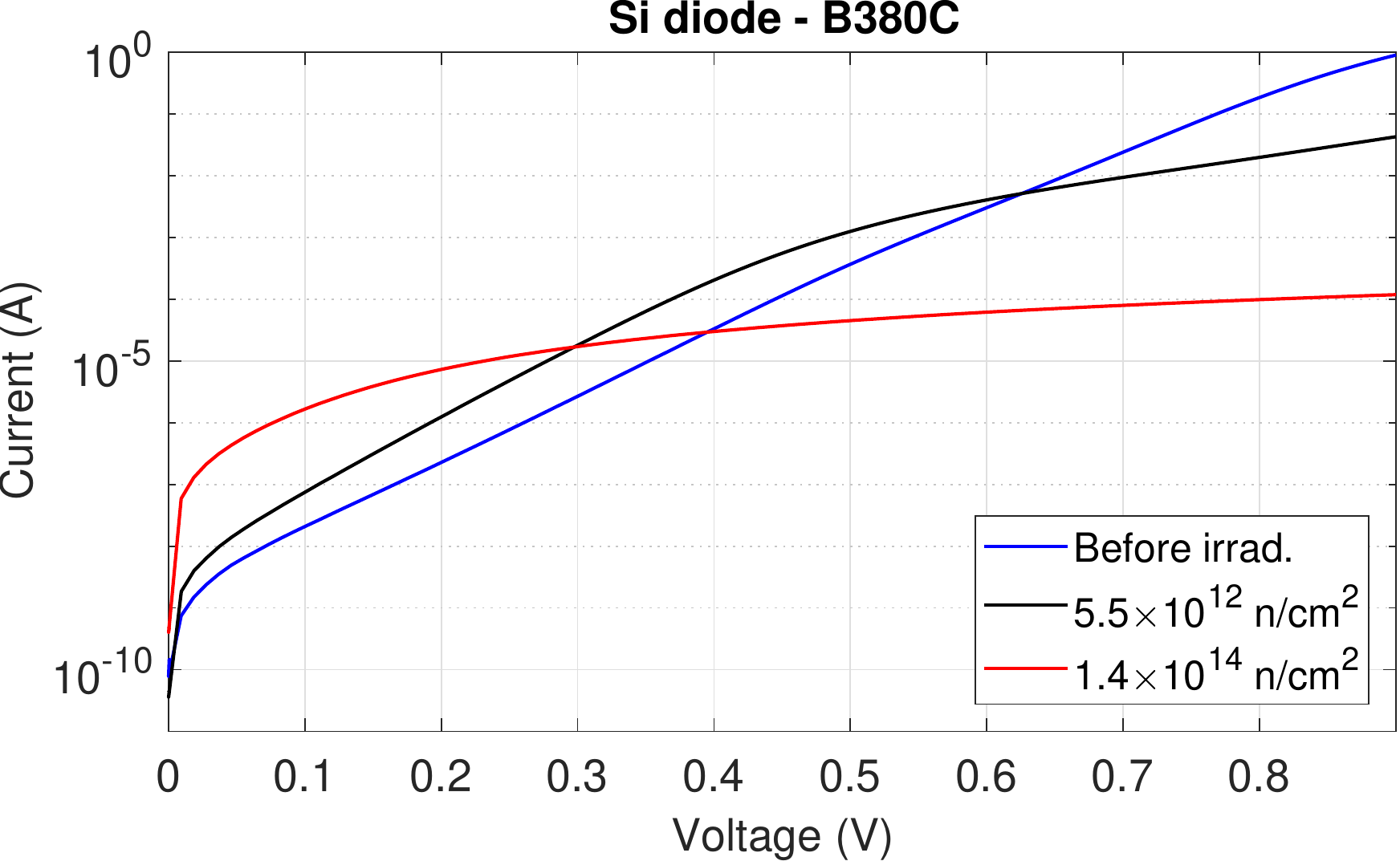}\label{fig:Si:forward}}\hfill
	\subfloat[]{\includegraphics[scale=0.45]{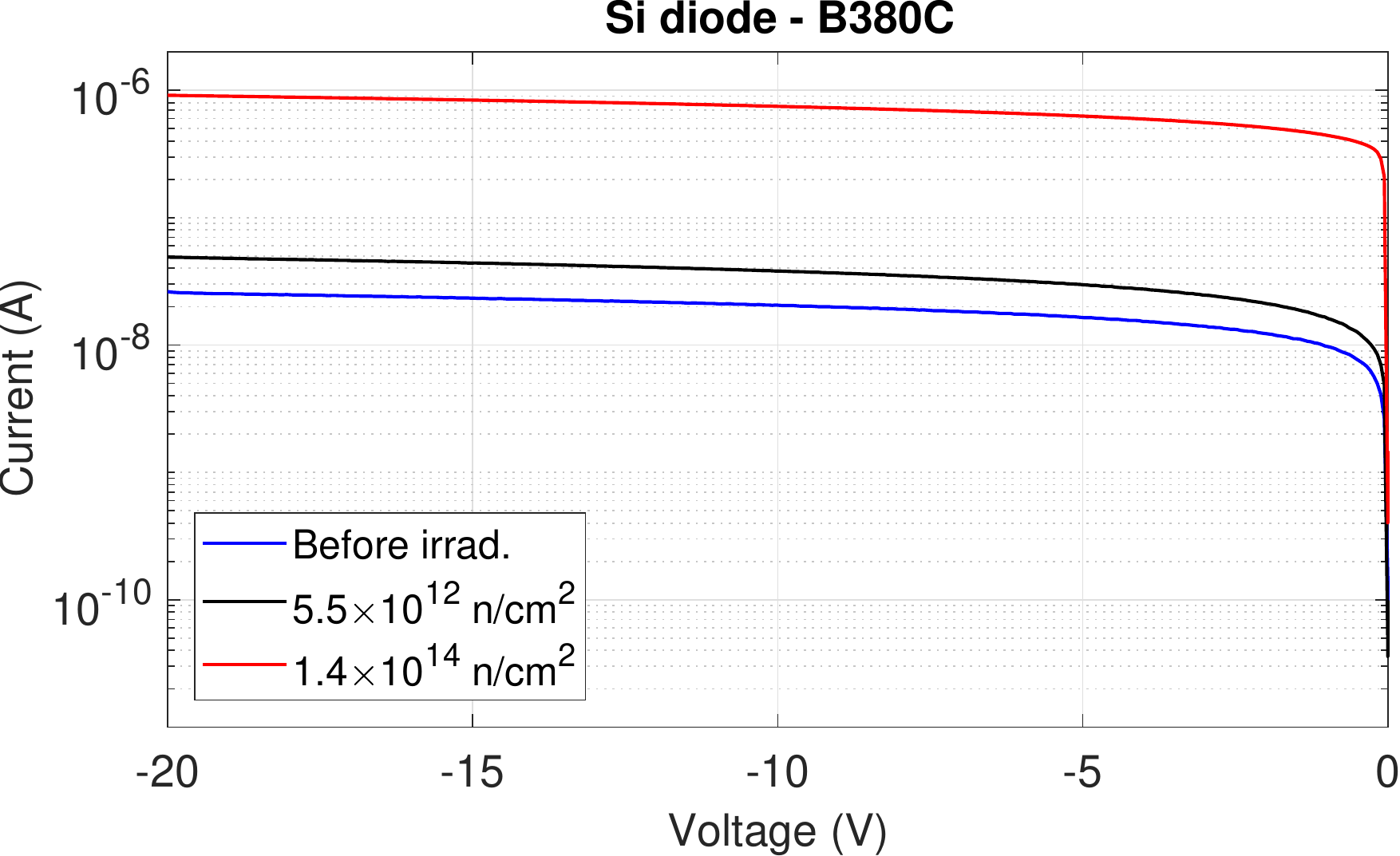}\label{fig:Si:reverse}}
	\caption{(a) Forward and (b) reverse \emph{I}--\emph{V} characteristics at room temperature of Si diodes of B380C bridge rectifiers, before and after irradiation (expressed as 1~MeV neq fluence).}
	\label{fig:Si}
\end{figure}

The forward and reverse \emph{I}--\emph{V} curves of B380C diodes before and after irradiation are illustrated in Figs. \ref{fig:Si:forward} and \ref{fig:Si:reverse}, respectively. The four diodes of each bridge exhibited similar characteristics, so only data of individual diodes is presented. 
High-energy proton irradiation introduces numerous defects in Si including vacancies, interstitials, divacancies, diinterstitials, vacancy-oxygen centres, C interstitials (C$_\text{i}$), complexes of O, C$_\text{i}$, H and Si vacancies and interstitials, as well as vacancy clusters \cite{davies2006radiation}. %vacanze, interstiziali, doppie vacanze, doppi interstiziali, vacanze-ossigeno (V-O), Ci, complessi tra O, Ci, H e vacanze/interstiziali, cluster di piu vacanze
These defects act as generation-recombination centres, traps or compensation centres, resulting in increased recombination current at low forward voltages, and increased series resistance due to carrier removal \cite{spieler1997introduction,bager2002lhcb,luo2003impact}. These effects are clearly observed in Fig. \ref{fig:Si:forward}. An increase of generation current at reverse bias is also measured: current at $-$20~V increases from 2.6$\times$10$^{-\text{8}}$~A before irradiation to 4.9$\times$10$^{-\text{8}}$~A after 5.5$\times$10$^\text{12}$~n/cm$^\text{2}$, and to 9.1$\times$10$^{-\text{7}}$~A after 1.4$\times$10$^\text{14}$~n/cm$^\text{2}$ (see Fig. \ref{fig:Si:reverse}). % come ovviamente previsto dalla letteratura.
%\textbf{(the reverse \emph{I}--\emph{V} curve will be included in the full paper)}. 
It is worth noting that Si diodes exhibit significant degradation at fluences more than an order of magnitude lower than SiC JBS diodes.

\section{Conclusions}
Transmission spectra of commercial-grade BS, FQ, PMMA, and PC samples irradiated with $\gamma$-rays up to 100~kGy are presented, and the degradation mechanism of each element is discussed. The obtained results indicate that, in terms of retained transparency upon irradiation, FQ is the most appropriate material for windows in rad-hard LED luminaires, while BS should be avoided; moreover, PMMA should be preferred over PC for secondary optics components. In addition, a Si diode bridge and a SiC JBS diode are tested using 24~GeV/\emph{c} protons, and radiation effects on both devices are discussed; the SiC JBS diodes measured were more than one order of magnitude more resistant than Si diodes, and therefore should be employed in future power supplies for rad-hard LED lights. In particular, the obtained results show that SiC JBS diodes and samples of FQ and PMMA exhibit no or limited degradation after radiation exposure levels equivalent to 10~years on the tunnel wall of a typical section of CERN accelerator complex, thus meeting the requirements on the expected lifetime of the rad-hard LED luminaires for CERN tunnels.  % nel senso che se anche la degradazione c'è, non pregiudica il funzionamento corretto del componente.

\section*{Acknowledgment}
The authors gratefully acknowledge Giuseppe Pezzullo (CERN) and Federico Ravotti (CERN) for performing sample irradiation and assistance at IRRAD facility, Elisa Guillermain (CERN) and Sophie Rouif (Ionisos) for managing the $\gamma$-ray irradiation at Ionisos, and Thomas Schneider (CERN) for providing the spectrophotometer and assistance at the Thin Film and Glass Laboratory (CERN).

\IEEEtriggeratref{36}


%\begin{thebibliography}{1}
%\end{thebibliography}

\bibliographystyle{IEEEtran}
\bibliography{RadHardLightingTNS_conf}

\begin{thebibliography}{10}
\providecommand{\url}[1]{#1}
\csname url@samestyle\endcsname
\providecommand{\newblock}{\relax}
\providecommand{\bibinfo}[2]{#2}
\providecommand{\BIBentrySTDinterwordspacing}{\spaceskip=0pt\relax}
\providecommand{\BIBentryALTinterwordstretchfactor}{4}
\providecommand{\BIBentryALTinterwordspacing}{\spaceskip=\fontdimen2\font plus
\BIBentryALTinterwordstretchfactor\fontdimen3\font minus
  \fontdimen4\font\relax}
\providecommand{\BIBforeignlanguage}[2]{{%
\expandafter\ifx\csname l@#1\endcsname\relax
\typeout{** WARNING: IEEEtran.bst: No hyphenation pattern has been}%
\typeout{** loaded for the language `#1'. Using the pattern for}%
\typeout{** the default language instead.}%
\else
\language=\csname l@#1\endcsname
\fi
#2}}
\providecommand{\BIBdecl}{\relax}
\BIBdecl

\bibitem{devine2016radiation}
J.~D. Devine and A.~Floriduz, ``Radiation hardening of {LED} luminaires for
  accelerator tunnels,'' in \emph{Proc. 16th Eur. Conf. Rad. Effects on
  Components and Systems (RADECS), \emph{Bremen (Germany)}}.\hskip 1em plus
  0.5em minus 0.4em\relax IEEE, 2016, pp. 1--6,
  \url{https://doi.org/10.1109/RADECS.2016.8093210}.

\bibitem{de2015radiation}
J.~P. De~Carvalho~Saraiva and M.~Brugger, ``Radiation environments and their
  impact at the {CERN}'s injector chain,'' CERN, ATS Note 2015-0042, Dec. 2015,
  \url{https://cds.cern.ch/record/2114889}.

\bibitem{floriduz2018modelling}
A.~Floriduz and J.~D. Devine, ``Modelling of proton irradiated {GaN}-based
  high-power white light-emitting diodes,'' \emph{Jpn. J. Appl. Phys.},
  vol.~57, no.~8, p. 080304, 2018,
  \url{https://doi.org/10.7567/JJAP.57.080304}.

\bibitem{bager2002lhcb}
T.~Bager, J.~Casas, B.~Palan, and M.~Rodriguez, ``Validation of switching power
  supplies, diode bridges, and conditioners for pressure sensors,'' CERN, EDMS
  document 1226409, 2002, \url{https://edms.cern.ch/document/1226409/1}.

\bibitem{b380c2013datasheet}
{Vishay General Semiconductor}, ``Glass passivated single-phase bridge
  rectifier,'' B380C Datasheet, [Revised Jul. 2013],
  \url{https://www.vishay.com/docs/88501/b40c1500g.pdf}.

\bibitem{gbu8k2015datasheet}
------, ``Glass passivated single-phase bridge rectifier,'' GBU8K Datasheet,
  [Revised Apr. 2015], \url{https://www.vishay.com/docs/88616/gbu8a.pdf}.

\bibitem{st2013sic}
{STMicroelectronics}, ``{New generation of 650 V SiC diodes},'' Application
  note AN4242, [Revised May 2013],
  \url{https://www.st.com/resource/en/application_note/dm00075656-new-generation-of-650-v-sic-diodes-stmicroelectronics.pdf}.

\bibitem{rouif2005radiation}
S.~Rouif, ``Radiation cross-linked polymers: Recent developments and new
  applications,'' \emph{Nucl. Instrum. Meth. Phys. Res. B}, vol. 236, no. 1-4,
  pp. 68--72, 2005, \url{https://doi.org/10.1016/j.nimb.2005.03.252}.

\bibitem{ravotti2015irrad}
F.~Ravotti, B.~Gkotse, M.~Moll, and M.~Glaser, ``{IRRAD}: The new 24 {GeV}/c
  proton irradiation facility at {CERN},'' in \emph{Proc. 12th Int. Topical
  Meeting Nucl. Applications of Accelerators (AccApp)}, vol.~15, 2015, pp.
  182--187, \url{http://cds.cern.ch/record/2237333}.

\bibitem{isidre2016niel}
I.~Mateu, M.~Moll, E.~Curras, F.~Ravotti, H.~Neugebauer, and M.~Glaser,
  ``{NIEL} hardness factor determination for the new proton irradiation
  facility at {CERN},'' CERN, Tech. Rep. AIDA-2020-SLIDE-2016-002, 2016,
  \url{https://cds.cern.ch/record/2162852}.

\bibitem{lebedev2000doping}
A.~Lebedev, A.~Veinger, D.~Davydov, V.~Kozlovski, N.~Savkina, and
  A.~Strel’chuk, ``Doping of n-type 6{H}--{SiC} and 4{H}--{SiC} with defects
  created with a proton beam,'' \emph{J. Appl. Phys.}, vol.~88, no.~11, pp.
  6265--6271, 2000, \url{https://doi.org/10.1063/1.1309055}.

\bibitem{doyle1998electrically}
J.~Doyle, M.~K. Linnarsson, P.~Pellegrino, N.~Keskitalo, B.~Svensson,
  A.~Sch{\"o}ner, N.~Nordell, and J.~Lindstr{\"o}m, ``Electrically active point
  defects in n-type 4{H}--{SiC},'' \emph{J. Appl. Phys.}, vol.~84, no.~3, pp.
  1354--1357, 1998, \url{https://doi.org/10.1063/1.368247}.

\bibitem{wang2012radiation}
Q.~Wang, Z.~Zhang, H.~Geng, C.~Sun, D.~Yang, S.~He, and Z.~Hu,
  ``Radiation-induced damage and recovery effects in {GG17} glass irradiated by
  1 {MeV} electrons,'' \emph{Nucl. Instrum. Meth. Phys. Res. B}, vol. 281, pp.
  1--7, 2012, \url{https://doi.org/10.1016/j.nimb.2012.04.010}.

\bibitem{wang2009effects}
Q.~Wang, H.~Geng, S.~He, D.~Yang, Z.~Zhang, X.~Qin, and Z.~Li, ``Effects of 80
  {keV} proton radiation on the optical properties and microstructure of
  type-{GG17} glass as rubidium lamp envelope,'' \emph{Nucl. Instrum. Meth.
  Phys. Res. B}, vol. 267, no.~15, pp. 2489--2494, 2009,
  \url{https://doi.org/10.1016/j.nimb.2009.06.006}.

\bibitem{wang2016radiation}
Q.~Wang, H.~Geng, C.~Sun, X.~Li, H.~Zhao, W.~Liu, J.~Xiao, and Z.~Hu,
  ``Radiation effects on optical and structural properties of {GG17} glasses
  induced by 170 {keV} electrons and protons,'' \emph{J. Appl. Phys.}, vol.
  119, no.~2, p. 023103, 2016, \url{ https://doi.org/10.1063/1.4939097}.

\bibitem{marshall1997induced}
C.~D. Marshall, J.~A. Speth, and S.~A. Payne, ``Induced optical absorption in
  gamma, neutron and ultraviolet irradiated fused quartz and silica,'' \emph{J.
  Non-Cryst. Solids}, vol. 212, no.~1, pp. 59--73, 1997,
  \url{https://doi.org/10.1016/S0022-3093(96)00606-0}.

\bibitem{leon2009gamma}
M.~Le{\'o}n, P.~Mart{\'\i}n, A.~Ibarra, and E.~Hodgson, ``Gamma irradiation
  induced defects in different types of fused silica,'' \emph{J. Nucl. Mater.},
  vol. 386, pp. 1034--1037, 2009,
  \url{https://doi.org/10.1016/j.jnucmat.2008.12.232}.

\bibitem{guzzi1993neutron}
M.~Guzzi, M.~Martini, A.~Paleari, F.~Pio, A.~Vedda, and C.~Azzoni, ``Neutron
  irradiation effects in amorphous $\text{SiO}_\text{2}$: optical absorption
  and electron paramagnetic resonance,'' \emph{J. Phys. Condens. Matter},
  vol.~5, no.~43, p. 8105, 1993,
  \url{https://doi.org/10.1088/0953-8984/5/43/022}.

\bibitem{mitchell1956cxi}
E.~Mitchell and E.~Paige, ``The optical effects of radiation induced atomic
  damage in quartz,'' \emph{Philos. Mag.}, vol.~1, no.~12, pp. 1085--1115,
  1956, \url{https://doi.org/10.1080/14786435608238193}.

\bibitem{paige1957kinetics}
E.~Paige, ``The kinetics of colour centre formation in quartz,'' \emph{Philos.
  Mag.}, vol.~2, no.~19, pp. 864--876, 1957,
  \url{https://doi.org/10.1080/14786435708242725}.

\bibitem{levy1955radiation}
M.~Levy and J.~Varley, ``Radiation induced colour centres in fused quartz,''
  \emph{Proc. Phys. Soc., Sect. B}, vol.~68, no.~4, p. 223, 1955,
  \url{https://doi.org/10.1088/0370-1301/68/4/304}.

\bibitem{rai2017uv}
V.~Rai, C.~Mukherjee, and B.~Jain, ``{UV-Vis and FTIR} spectroscopy of gamma
  irradiated polymethyl methacrylate,'' \emph{Indian J. Pure Appl. Phys.},
  vol.~55, no.~11, pp. 775--785, 2017,
  \url{http://op.niscair.res.in/index.php/IJPAP/article/view/15541}.

\bibitem{clough1996color}
R.~Clough, K.~Gillen, G.~Malone, and J.~Wallace, ``Color formation in
  irradiated polymers,'' \emph{Radiat. Phys. Chem.}, vol.~48, no.~5, pp.
  583--594, 1996, \url{https://doi.org/10.1016/0969-806X(96)00075-8}.

\bibitem{lu2000transmittance}
K.-P. Lu, S.~Lee, and C.~P. Cheng, ``Transmittance in irradiated poly (methyl
  methacrylate) at elevated temperatures,'' \emph{J. Appl. Phys.}, vol.~88,
  no.~9, pp. 5022--5027, 2000, \url{https://doi.org/10.1063/1.1316792}.

\bibitem{sharma2008modification}
T.~Sharma, S.~Aggarwal, A.~Sharma, S.~Kumar, V.~Mittal, P.~Kalsi, and
  V.~Manchanda, ``Modification of optical properties of polycarbonate by gamma
  irradiation,'' \emph{Radiat. Eff. Defects Solids}, vol. 163, no.~2, pp.
  161--167, 2008, \url{https://doi.org/10.1080/10420150701688503}.

\bibitem{sinha2004gamma}
D.~Sinha, K.~Sahoo, U.~Sinha, T.~Swu, A.~Chemseddine, and D.~Fink,
  ``Gamma-induced modifications of polycarbonate polymer,'' \emph{Radiat. Eff.
  Defects Solids}, vol. 159, no.~10, pp. 587--595, 2004,
  \url{https://doi.org/10.1080/10420150412331330539}.

\bibitem{factor1994chemistry}
A.~Factor, J.~Carnahan, S.~Dorn, and P.~Van~Dort, ``The chemistry of
  $\gamma$-irradiated bisphenol-{A} polycarbonate,'' \emph{Polym. Degrad.
  Stab.}, vol.~45, no.~1, pp. 127--137, 1994,
  \url{https://doi.org/10.1016/0141-3910(94)90188-0}.

\bibitem{factor1997use}
A.~Factor and P.~Donahue, ``The use of $^{31}${P} {NMR} to identify color
  bodies in $\gamma$-irradiated bisphenol-{A} polycarbonate,'' \emph{Polym.
  Degrad. Stab.}, vol.~57, no.~1, pp. 83--86, 1997,
  \url{https://doi.org/10.1016/S0141-3910(96)00216-9}.

\bibitem{gusarov2005comparison}
A.~Gusarov, D.~Doyle, L.~Glebov, and F.~Berghmans, ``Comparison of
  radiation-induced transmission degradation of borosilicate crown optical
  glass from four different manufacturers,'' in \emph{Proc. SPIE 5897,
  Photonics for Space Environments X}.\hskip 1em plus 0.5em minus 0.4em\relax
  International Society for Optics and Photonics, 2005, p. 58970I,
  \url{https://doi.org/10.1117/12.619199}.

\bibitem{agnello2006thermal}
S.~Agnello and L.~Nuccio, ``Thermal stability of gamma-irradiation-induced
  oxygen-deficient centers in silica,'' \emph{Phys. Rev. B}, vol.~73, no.~11,
  p. 115203, 2006, \url{https://doi.org/10.1103/PhysRevB.73.115203}.

\bibitem{clough1995discoloration}
R.~Clough, G.~Malone, K.~Gillen, J.~Wallace, and M.~Sinclair, ``Discoloration
  and subsequent recovery of optical polymers exposed to ionizing radiation,''
  \emph{Polym. Degrad. Stab.}, vol.~49, no.~2, pp. 305--313, 1995,
  \url{https://doi.org/10.1016/0141-3910(95)87013-X}.

\bibitem{treadaway1975transient}
M.~J. Treadaway, B.~C. Passenheim, and B.~D. Kitterer, ``Transient radiation
  effects in optical materials,'' Intelcom Rad Tech, Tech. Rep.
  SAMSO-TR-75-174, 1975,
  \url{https://apps.dtic.mil/dtic/tr/fulltext/u2/a013922.pdf}.

\bibitem{treadaway1976radiation}
M.~Treadaway, B.~Passenheim, B.~Kitterer, and P.~Schall, ``Radiation coloration
  and bleaching of glass,'' \emph{IEEE Trans. Nucl. Sci.}, vol.~23, no.~6, pp.
  1820--1825, 1976, \url{https://doi.org/10.1109/TNS.1976.4328584}.

\bibitem{lundy34methods}
C.~Lundy, M.~Licata, E.~Haag, and S.~Krishnan, ``Methods of color stabilization
  in radiation-sterilized polycarbonate,'' in \emph{Proc. SPE ANTEC
  1988}.\hskip 1em plus 0.5em minus 0.4em\relax Society of Plastics Engineers,
  1988, pp. 1348--1351.

\bibitem{hermanson1993physical}
N.~Hermanson and J.~Steffen, ``The physical and visual property changes in
  thermoplastic resins after exposure to high energy sterilization-gamma versus
  electron beam,'' in \emph{Proc. SPE ANTEC 1993}.\hskip 1em plus 0.5em minus
  0.4em\relax Society of Plastics Engineers, 1993, pp. 784--784.

\bibitem{chadwick1972effect}
K.~Chadwick, ``The effect of light exposure on the optical density of
  irradiated clear polymethylmethacrylate,'' \emph{Phys. Med. Biol.}, vol.~17,
  no.~1, pp. 88--93, 1972, \url{https://doi.org/10.1088/0031-9155/17/1/010}.

\bibitem{harris2005displacement}
R.~D. Harris, A.~J. Frasca, and M.~O. Patton, ``Displacement damage effects on
  the forward bias characteristics of {SiC Schottky} barrier power diodes,''
  \emph{IEEE Trans. Nucl. Sci.}, vol.~52, no.~6, pp. 2408--2412, 2005,
  \url{https://doi.org/10.1109/TNS.2005.860730}.

\bibitem{nava2008silicon}
F.~Nava, G.~Bertuccio, A.~Cavallini, and E.~Vittone, ``Silicon carbide and its
  use as a radiation detector material,'' \emph{Meas. Sci. Technol.}, vol.~19,
  no.~10, p. 102001, 2008,
  \url{https://doi.org/10.1088/0957-0233/19/10/102001}.

\bibitem{luo2003impact}
Z.~Luo, T.~Chen, J.~D. Cressler, D.~C. Sheridan, J.~R. Williams, R.~A. Reed,
  and P.~W. Marshall, ``Impact of proton irradiation on the static and dynamic
  characteristics of high-voltage {4H-SiC JBS} switching diodes,'' \emph{IEEE
  Trans. Nucl. Sci.}, vol.~50, no.~6, pp. 1821--1826, 2003,
  \url{https://doi.org/10.1109/TNS.2003.821806}.

\bibitem{kozlovski2018electrical}
V.~Kozlovski, A.~Lebedev, M.~Levinshtein, S.~Rumyantsev, and J.~Palmour,
  ``{Electrical and noise properties of proton irradiated 4H-SiC Schottky
  diodes},'' \emph{J. Appl. Phys.}, vol. 123, no.~2, p. 024502, 2018,
  \url{https://doi.org/10.1063/1.5018043}.

\bibitem{messenger2003niel}
S.~R. Messenger, E.~A. Burke, M.~A. Xapsos, G.~P. Summers, R.~J. Walters,
  I.~Jun, and T.~Jordan, ``{NIEL} for heavy ions: An analytical approach,''
  \emph{IEEE Trans. Nucl. Sci.}, vol.~50, no.~6, pp. 1919--1923, 2003,
  \url{https://doi.org/10.1109/TNS.2003.820762}.

\bibitem{lee2003comparative}
K.~Lee, T.~Ohshima, A.~Saint, T.~Kamiya, D.~Jamieson, and H.~Itoh, ``A
  comparative study of the radiation hardness of silicon carbide using light
  ions,'' \emph{Nucl. Instrum. Meth. Phys. Res. B}, vol. 210, pp. 489--494,
  2003, \url{https://doi.org/10.1016/S0168-583X(03)01096-6}.

\bibitem{khanna2004proton}
S.~M. Khanna, D.~Estan, L.~S. Erhardt, A.~Houdayer, C.~Carlone,
  A.~Ionascut-Nedelcescu, S.~R. Messenger, R.~J. Walters, G.~P. Summers, J.~H.
  Warner, and I.~Jun, ``Proton energy dependence of the light output in gallium
  nitride light-emitting diodes,'' \emph{IEEE Trans. Nucl. Sci.}, vol.~51,
  no.~5, pp. 2729--2735, 2004, \url{https://doi.org/10.1109/TNS.2004.835097}.

\bibitem{davies2006radiation}
G.~Davies, S.~Hayama, L.~Murin, R.~Krause-Rehberg, V.~Bondarenko, A.~Sengupta,
  C.~Davia, and A.~Karpenko, ``Radiation damage in silicon exposed to
  high-energy protons,'' \emph{Phys. Rev. B}, vol.~73, no.~16, p. 165202, 2006,
  \url{https://doi.org/10.1103/PhysRevB.73.165202}.

\bibitem{spieler1997introduction}
H.~Spieler, ``Introduction to radiation-resistant semiconductor devices and
  circuits,'' in \emph{AIP Conf. Proc.}, vol. 390, no.~1.\hskip 1em plus 0.5em
  minus 0.4em\relax AIP, 1997, pp. 23--49,
  \url{https://doi.org/10.1063/1.52282}.

\end{thebibliography}

\end{document}